\documentclass[12pt, preprint]{aastex}
\slugcomment{Eccentricity Evolution of Extrasolar Planets}
\shorttitle{ECCENTRICITY EVOLUTION OF EXTRASOLAR PLANETS}
\shortauthors{NAGASAWA, LIN, \& IDA}

\begin{document}
\title{Eccentricity Evolution of Extrasolar Multiple Planetary
Systems due to the Depletion of Nascent Protostellar Disks}

\author{M. Nagasawa\altaffilmark{1,2}, D. N. C. Lin\altaffilmark{1}, and
S. Ida\altaffilmark{3}}
\altaffiltext{1}{UCO/Lick Observatory, University of
California, Santa Cruz, CA 95064, U.S.A.}
\altaffiltext{2}{Present address: Space Science Division, MS 245-3,
NASA Ames Research Center, Moffett Field, CA 94035-1000, U.S.A.}
\altaffiltext{3}{Department of Earth and Planetary Sciences,
Tokyo Institute of Technology, Meguro-ku, Tokyo 152-8551,
Japan}
\email{mnagasawa@mail.arc.nasa.gov, lin@ucolick.org, ida@geo.titech.ac.jp}

\begin{abstract}
Most extrasolar planets are observed to have eccentricities much
larger than those in the solar system. Some of these planets have
sibling planets, with comparable masses, orbiting around the same host
stars.  In these multiple planetary systems, eccentricity is modulated
by the planets' mutual secular interaction as a consequence of angular
momentum exchange between them. For mature planets, the
eigenfrequencies of this modulation are determined by their mass and
semi-major axis ratios.  But, prior to the disk depletion, self gravity
of the planets' nascent disks dominates the precession
eigenfrequencies.  We examine here the initial evolution of young
planets' eccentricity due to the apsidal libration or circulation
induced by both the secular interaction between them and the self
gravity of their nascent disks.  We show that as the latter effect
declines adiabatically with disk depletion, the modulation amplitude
of the planets' relative phase of periapse is approximately invariant
despite the time-asymmetrical exchange of angular momentum between
planets.  However, as the young planets' orbits pass through a state
of secular resonance, their mean eccentricities undergo systematic
quantitative changes.  For applications, we analyze the eccentricity
evolution of planets around Upsilon Andromedae and HD168443 during the
epoch of protostellar disk depletion. We find that the disk depletion
can change the planets' eccentricity ratio.  However, the relatively
large amplitude of the planets' eccentricity cannot be excited if all
the planets had small initial eccentricities.
\end{abstract}


\section{INTRODUCTION}

Many extrasolar planetary systems have been discovered recently using
the radial velocity technique (Marcy \& Butler 2000).  The basic
assumption is that the spectroscopic variations observed in some
targeted stars are due to the Doppler shift associated with the reflex
motion of stars with unseen companions.  The mass of these companions
depends on the poorly known inclination of these systems.  But, unless
the orbits of these systems are highly inclined (Stepinski \& Black
2001), the inferred masses of the companions are comparable to that of
Jupiter. The observed spectra of some targeted stars indicate the
presence of multiple companions around them.  For example, three
companions are found to orbit around Upsilon Andromedae. The long term
stability of this system requires their inclination to be sufficiently
small such that the masses of these companions are no more than a few
Jupiter mass (${M_J}$) and much smaller than that of their host star
(Laughlin \& Adams 1999; Rivera \& Lissauer 2000; Stepinski, Malhotra,
\& Black 2000; Ito \& Miyama 2001; Lissauer \& Rivera 2001).  Two
companions are also found around HD168443 with minimum masses which
are an order of magnitude larger than ${M_J}$.  Although the masses of
these companions would be a modest fraction of a solar mass
($M_\odot$) if this system is viewed nearly face on, such compact
hierarchical stellar systems have not be seen before.  Thus, we follow
the conventional practice to refer these multiple companions as
planets.

In the limit that the planets' masses are substantially smaller than
that of their host star, their mutual secular perturbation induces
them to exchange angular momentum while preserving their energy.
(Planets in mean motion resonances also exchange energy on comparable
time scales).  Around Ups And and HD168443, the planets' orbits are
not in mean motion resonances so that their semi-major axes are
conserved while their eccentricity and longitude of periapse modulate
over some characteristic secular time scale (Murray \& Dermott 1999).

But, the secular perturbation between the planets has not always been
sustained at the present level.  During the epoch of their formation,
protoplanets are embedded in protostellar disks which have been found
around most young stellar objects (cf. Haisch, Lada, \& Lada 2001).
The mass and temperature distribution of these disks are very similar
to those inferred from the minimum mass nebula model for the solar
system (Beckwith 1999).  The self gravity of these disks can induce
the orbits of planets formed within them to precess at a rate faster
than that due their mutual perturbation. Consequently, the rate of
angular momentum transfer between the interacting planets is
suppressed. 

In the solar system, the initial contribution of the solar nebula to
the total gravitational potential dominates the precession frequency
of asteroids and comets over that due to the secular perturbation
induced on them by the giant planets.  The planets also undergo
precession induced by the disk gravity and other planets' secular
perturbation.  In general the precession frequencies of these
celestial bodies do not equal each other.  But as gas is depleted in
the solar nebula, its self gravity weakens.  The total precession
frequencies of both the asteroids and the planets declines, though at
a different rate. When the precession frequency of asteroids with some
semi major axes coincides with that of a major planet, they enter into
a state of secular resonance.  In this resonance, the eccentricity of
the asteroids is either excited or damped monotonically, depending on
their relative longitude of periastron passage with respect to that of
the planet.  As this secular resonance sweeps across the solar system,
large eccentricity may be excited among some small celestial bodies
(Ward, Colombo, \& Franklin 1976; Heppenheimer 1980; Nagasawa, Tanaka,
\& Ida 2000; Nagasawa \& Ida 2000).

In this paper, we examine the effects of disk depletion on the
eccentricity evolution of the planetary systems around Ups And and
HD168443.  Through such an investigation, we hope to infer the
kinematic properties which these planets are born with and thereby
cast constraints on their formation process. In \S2, we briefly
discuss the precession due to the secular interaction between planets
and that due to the self gravity of the protostellar disk. The
conditions for secular resonance are discussed.  We introduce, in \S3,
a working model and describe the method we used to analyze the
eccentricity evolution.  In \S4, we present the results of some
calculations. We use these numerical results to demonstrate the
evolution of these systems with Hamiltonian contour maps.  Based on
these results, we infer, in \S5, some implications on the planets'
orbital eccentricity shortly after their formation and while they are
still embedded in their nascent disks.  Finally, we summarize our
results in \S6.

\section{SECULAR INTERACTION IN MULTIPLE EXTRASOLAR PLANETARY SYSTEMS}

\subsection{Current Orbital Properties}

In this paper, we focus our discussion on the planets around Upsilon
Andromeda (c and d) and those around HD168443.  The decomposition of
Upsilon Andromeda's spectra indicate that there are three planets
orbiting around it (Butler et al. 1999). Inferred orbital
elements\footnote{http://exoplanets.org/almamacframe.html} of the
Upsilon Andromeda planets are shown in Table \ref{tab:parameta}(a). 
The semi-major axis, the eccentricity, the longitude of periastron,
mass of the planet, and periastron passage time (JD) are denoted by
$a$, $e$, $\varpi$, $M$, and $T_{\rm peri}$, respectively. The
subscripts $b, c,$ and $d$ represent the values for the individual
planets.  Throughout this analysis, we assume that these systems are
viewed edge-on and their masses correspond to their minimum values.
Provided their orbits are coplanar, our analysis is independent of the
planets' inclination.  Our approach becomes inadequate for the
limiting cases of nearly face-on orbits where the masses of the
companions are comparable to their host stars.

Around Ups And, the eccentricity of the innermost planet b is
essentially undetectable.  The orbit of planet b is likely to be
circularized during the main sequence life span of the host star Ups
And by the tidal dissipation within the planet's interior as expected
for Jupiter-like extrasolar planets with semi-major axis less than
0.05 AU (Rasio et al. 1996).  With its low mass and small semi-major
axis, planet b does not contribute significantly to the dynamical
evolution of the system (Mardling \& Lin 2003).  Although the outer
planets (c and d) still exert secular perturbation on planet b, the
cumulative effect on its eccentricity modulation is limited.  Under
the present configuration, the secular interaction of planet b with
planets c and d is weakened by the rapid precession due to the
post-Newtonian relativistic correction in the gravitational potential
of the host star (Mardling \& Lin 2003).  Neglecting the secular
perturbation due to planet b, the orbital evolution of planets c and d
obtained from the direct orbital integration of the full equation of
motion is shown in Figure \ref{fig:orbitnow}a. In the numerical
integration, the calculation is started with the eccentricity ratio
$x\equiv e_c/e_d$ and the relative longitude of periastrons between
planets c and d, $\eta \equiv \varpi_c - \varpi_d$, based on their
observed values $0.66$ and $\sim -0.06$ radian, respectively. We
compute the evolution of the system for over $2\times10^4$ years. The
equi-Hamiltonian contours (see Appendix \ref{sec:hamil}) are also
shown.  For planets c and d around Ups And, the massive planet d has
larger eccentricity than smaller planet c.  The longitudes of
periastron of planet c and d are always close to each other.  When
the planets are in librating (closed) track in ($x$-$\eta$) diagram,
the planetary system tends to be stable for a long time. The
stabilities of this system around Ups And are well investigated
(Laughlin \& Adams 1999; Rivera \& Lissauer 2000; Stepinski et
al. 2000; Barnes \& Quinn 2001; Ito \& Miyama 2001; Lissauer \& Rivera
2001; Chiang, Tabachnik, \& Tremaine 2002; Mardling \& Lin 2003).

Around HD168443, two massive planets are inferred from the radial
velocity curves (Marcy et al. 2001). The best-fit orbital elements
are shown in Table \ref{tab:parameta}(b).  We also numerically
integrate the present-day orbital evolution of the planets b and c
around HD168443.  In contrast to the planetary system around Ups And,
the lower mass planet b around HD168443 system, has an eccentricity
which is more than twice that of the massive planet c. The orbits are
integrated over $2.5\times10^4$ years, starting with $x =2.65$ and
$\eta \sim 1.92$ radian. These planets evolve in circulating (open)
track in the ($x$-$\eta$) diagram (Fig. \ref{fig:orbitnow}b) and
$\eta$ evolves from $-\pi$ to $\pi$ over a period of about $1.6 \times
10^4$ years.  This system too is stable despite the large magnitude of
eccentricity of planet b.  The origin of such large mass and
eccentricity of planets in this system has not been addressed
previously.

\subsection{Secular Perturbation between Two Planets}

The secular interaction induces eccentricity modulation between
planets c and d around Ups And (Laughlin \& Adams 1999; Rivera \&
Lissauer 2000; Ito \& Miyama 2001; Mardling \& Lin 2003).  For
presentation purpose, it is useful to briefly recapitulate the
analysis of secular interaction between two planets.  It is customary
to consider the secular (long-term) evolution of the interacting
planets' orbits using a disturbing function (Murray \& Dermott 1999).
To the {\it lowest order}, the modulation of the eccentricity
($e_{c,d}$) and longitude of periapse ($\varpi_{c,d}$) of some planets
c and d can be approximated by
\begin{equation}
{d e_{c,d} \over d \tau} =\gamma_{c,d} C e_{d,c} {\rm sin}
(\varpi_{c,d} - \varpi_{d,c}) + \Lambda_{c,d} ^e,
\label{eq:dedtau}
\end{equation}
\begin{equation}
{d \varpi_{c,d} \over d \tau} = \gamma_{c,d} \left( 1 +
C \left( {e_{d,c} \over e_{c,d}
} \right) {\rm cos} (\varpi_{c,d} - \varpi_{d,c}) \right)+ \Lambda_{c,d}
^\varpi + N_{c,d},
\label{eq:dwdtau}
\end{equation}
where $\tau \equiv t/t_c$, $t$ is the real time, $t_c = (4 / n_c)
(M_\ast / M_d) (a_d/a_c)^2/ b_{3/2}^{(1)}$, $b_{s}^{(j)}(\alpha)$ are
the Laplace coefficients with $\alpha=a_c/a_d$, $M_c$, $M_d$, and
$M_\ast$ are the mass of planets c, d, and the host star respectively,
$a_c$ and $a_d$ are the semi-major axes of planets c and d
respectively, $n_{c,d} = (GM_\ast/a_{c,d}^3)^{1/2}$ is the mean motion
of planet c or d, $\gamma_{c,d} \equiv (a_c/a_{c,d})^{1/2} (M_c/ M_{c,d})$
so that $\gamma_c=1$, $C \equiv -b_{3/2}^{(2)}/b_{3/2}^{(1)}$ (Mardling
\& Lin 2002), $\Lambda_{c,d} ^e$ is the change rate of eccentricity due to
the non axisymmetric torque, $\Lambda_{c,d} ^\varpi$ is the change rate of
$\varpi$ due to the non axisymmetric torque, and $N_{c,d}$ is
axisymmetric modification of the gravitational potential.  When
$a_c/a_d \ll 1$, $t_c=(4 /3 n_c)(M_\ast / M_d) (a_d/a_c)^3$ and $C=- 5
a_c/4 a_d$.  Frequently used variables are tabulated in
Table \ref{tab:variables}.  The planets' secular interaction induces
both axisymmetric and non axisymmetric perturbations on each other.
While only the non axisymmetric component of the planets' secular
perturbation on each other can lead to angular momentum exchange and
$e_{c,d}$ modulations in equation (\ref{eq:dedtau}), both axisymmetric
and non axisymmetric perturbations contribute to the evolution of
$\varpi_{c,d}$ through the first and second term on the right hand
side of equation ({\ref{eq:dwdtau}) respectively. These linearized
equations of secular motion, though only accurate to the first order,
are useful for obtaining analytic-approximation solutions which
highlight the dominant physical effects.  For the numerical
calculation of the planets' orbits (see \S4), we use the exact
equation of planets' motion.

Planet-disk interaction also leads to both axisymmetric modification
of the gravitational potential $(N_{c,d})$ and non axisymmetric torque
$(\Lambda_{c,d} ^{e, \varpi})$ (Goldreich \& Tremaine 1980, 1982; Lin
\& Papaloizou 1986 a, b, 1993).  For the latter effect, planets excite
waves in the disk which carry angular momentum.  The dissipation of
these waves, anywhere in the disk, would result in a finite torque
(Papaloizou \& Lin 1984).  In principle, the net torque vanishes in
the invisid limit.  However, their amplitude grows and steepens into
nonlinear shocks as the waves propagate away from the location where
they are launched, leading to an effective torque (Savonije,
Papaloizou, \& Lin 1994).  The evolution of the axisymmetric potential
modifies the precession of $\varpi$ but do not directly influence the
eccentricity of the orbits (see eq. [\ref{eq:dwdtau}]).  While the non
axisymmetric torque may induce monotonic changes in $e$ (Chiang \&
Murray 2002) and $\varpi$ (see eq. [\ref{eq:dedtau}]), the magnitude and
sign of $\Lambda^e _{c,d}$ depends sensitively on the disk structure.
Interaction between the embedded planets with disk gas through
corotation resonances damps the eccentricity while that through
Lindblad resonances excites the eccentricity (Goldreich \& Tremaine
1980).  For protoplanets with mass less than a few times that of
Jupiter and modest eccentricity, gas may flow in the vicinity of their
orbits such that the eccentricity damping effect of the corotation
resonances is stronger than the excitation effect of the Lindblad
resonances (Goldreich \& Tremaine 1980).  But protoplanets with masses
an order of magnitude larger than that of Jupiter may open relatively
wide gaps in protostellar disks.  In this limit, the protoplanets'
corotation resonances may be cleared of disk gas such that their
eccentricity may be excited (Artymowicz 1993; Papaloizou, Nelson, \&
Masset 2001; Goldreich \& Sari 2002). For precession, the contribution
from the non axisymmetric torque may be weaker than that due to the
disk's axisymmetric contribution to the total gravitational potential
in the limit that the torque resulting from the planets' interaction
with the interior and exterior regions of the disk are balanced or in
low-viscosity and thin disks where planets with relatively low masses
can open wide gaps.  However, the non axisymmetric torque may lead to
significant contributions over time in relatively massive disks.

In order to analyze the contribution of each effect, we adopt a piece
meal approach.  In this paper, we focus our attention on the evolution
of planetary orbits due to the changes in the axisymmetric disk
potential.  For the departure from a point-mass potential, the apsidal
motion of planets c and d are included in $N_{c,d}$ (see Appendix
\ref{sec:disk}).  In this first step, we neglect the effects due to
the non axisymmetric torque by setting $\Lambda_{c,d} ^{e, \varpi}=0$
in equations (\ref{eq:dedtau}) and ({\ref{eq:dwdtau}). The extension of our
discussion to the limit of finite non axisymmetric torque will be
presented in a future contribution.

\subsection{Precession Frequencies}

To the lowest order, the solution of equations (\ref{eq:dedtau}) and
(\ref{eq:dwdtau}) can be expressed as
\begin{equation}
e_{c,d} \exp (i \varpi_{c,d}) = A_{c,d} \exp \{i (g_1 t + \beta_1)\}
+ B_{c,d} \exp \{i (g_2 t + \beta_2)\},
\label{eq:ecdvar}
\end{equation}
where $A_{c,d}$ and $B_{c,d}$ are the oscillation amplitudes,
$\beta_1$ and $\beta_2$ are the phase angles (e.g., Brouwer \&
Clemence 1961). The individual planet's longitudes of periastrons
precess with two independent eigenfrequencies,
\begin{equation}
g_{1,2}=\frac{1}{2}\left\{ g_c+g_d \pm \left[ (g_c-g_d)^2+4 g_{cd}^2
\right]^{1/2}  \right\},
\label{eq:twoeign}
\end{equation}
where
\begin{equation}
g_{c,d} \equiv {\gamma_{c,d} + N_{c,d} \over t_c}, 
\ \ g_{cd} \equiv {C \gamma_d^{1/2}
\over t_c} \propto (M_c M_d)^{1/2}.
\label{eq:gcandd}
\end{equation}
From equation (\ref{eq:dwdtau}), we find that $g_c$ would be the precession
frequency of planet c if it is massless and is subjected to the
secular perturbation of a massive planet d with a circular orbit.
Similarly, $g_d$ would be the precession frequency of planet d if it
is massless and is subjected to the secular perturbation of a massive
planet c with a circular orbit.

In the limit that $A_{c,d}$ and $B_{c,d}$ have comparable magnitudes
and the two eigenfrequencies, $g_{1,2}$ have sufficiently large
differences, the relative longitude of the two planets would
circulate.  Precessional degeneracy occurs when $g_1 = g_2$ so that the
two planets precess synchronously regardless of the amplitude of their
$e$ and $\varpi$.  In this degenerate state, the relative phase of the
two planets' longitude of periapse passage, $\eta \equiv \varpi_c - 
\varpi_d$, retains a constant
value and the two planets are in a state of secular resonance.  In
this secular resonance, angular momentum is monotonically transferred
from one planet to another because the relative orientation of their
periastrons is maintained indefinitely. To lowest order, the planets'
individual energy is conserved.  With the exception of the special
configuration in which $\eta = 0$ or $\pm \pi$, this
monotonic transfer generally leads to eccentricity excitation of the
angular momentum losing bodies and eccentricity damping of the angular
momentum gaining bodies.  In celestial mechanics, secular resonance
is thought to be important in the current eccentricity distribution of
the asteroids (Williams 1969).

From equation (\ref{eq:twoeign}), we find that the magnitude of the
difference between $g_1$ and $g_2$ is
\begin{equation}
\Delta g_{12} \equiv \vert g_1 - g_2 \vert = \left[ (g_c-g_d)^2+4 g_{cd}
\right]^{1/2} = \left[ \left( { 1 - \gamma_d + \Delta N \over
\tau_c } \right)^2+4 g_{cd} \right]^{1/2},
\label{eq:dg12}
\end{equation}
which vanishes on the exact center of secular resonance (where
$g_1=g_2$) only if $g_c=g_d$ and $g_{cd} =0$. The former requirement
corresponds to a necessary resonance condition $A_1 = 0$ where
\begin{equation}
A_1 \equiv 1 - \gamma_d + \Delta N
\label{eq:rescon}
\end{equation}
is the precession rate induced by axisymmetric component
of the perturbed potential (see eq. [\ref{eq:detadt}] below).  The latter
requirement is satisfied if either $M_c=0$ or $M_d=0$ (see eq. 
[\ref{eq:gcandd}]).  Equation (\ref{eq:dg12}) implies that planets with
comparable masses cannot have precessional degeneracy with $g_1 = g_2$
(Kinoshita \& Nakai 2000).

The present value of $\gamma_d $ is 0.29 for planets c and d around Ups
And.  Today, in the absence of any residual disk ($N_{c,d} = 0$), $g_2
\sim g_d < g_1 \sim g_c$ and the outer two planets of Ups And are {\it
not} in a state of precessional degeneracy.  Nevertheless, their
present orbits can be approximated by equation (\ref{eq:ecdvar}) with the
magnitude of $A_{c,d}$ much smaller than that of $B_{c,d}$.  Thus, the
two planets primarily precess with eigenfrequency $g_1$ and their
relative longitudes of periastrons librate over a restricted range of
phases.  This phase lock is equivalent of a state of secular
resonance.  This resonant interaction is the result of a dynamical
feedback through the planets' secular interaction.

\subsection{Three Classes of Relative Orbits}

In order to illustrate the importance of this feedback effect, we
substitute $x \equiv e_c/e_d$ and $\eta \equiv \varpi_c - \varpi_d$,
equations (\ref{eq:dedtau}) and (\ref{eq:dwdtau}) reduce to
\begin{equation}
{d x \over d \tau} = C (1 + \gamma_d x^2) {\rm sin} \eta,
\label{eq:dxdtau}
\end{equation}
\begin{equation}
{d \eta \over d \tau}
= (1 - \gamma_d)+(C/x) ( 1 - \gamma_d x^2 ) {\rm cos} \eta + \Delta N
=A_1 + A_2 (x) {\rm cos} \eta,
\label{eq:detadt}
\end{equation}
where $\Delta N \equiv N_c - N_d$, $A_1$ is given in
equation (\ref{eq:rescon}), and $A_2 (x) \equiv (C/x) ( 1 - \gamma_d x^2 )$
is introduce for notational simplicity.  There are three families of
relative orbits which can be illustrated below with the linearized
approximation solutions of equations (\ref{eq:dxdtau}) and (\ref{eq:detadt})
(see Appendix \ref{sec:libra}).  We show below that the relative
magnitude of $A_1$ and $A_2$ determine the nature of the orbits.

1) {\it Circulation.}  In these solutions, the magnitude of $x$
modulates about some values $x_0$ such that $x = x_0 + \delta x(\tau)$
with an amplitude $\vert \delta x(\tau) \vert \ll  x_0$ for all
values of $\eta$ which ranges between 0 and $2 \pi$.  In the limit
that $\vert A_1 \vert \gg \vert A_2 \vert$, $\eta$ decreases
monotonically (because $A_1 < 0$) while $\delta x$ oscillates (see
solutions in eqs. [\ref{eq:eta1b}] and [\ref{eq:x1}] in Appendix
\ref{sec:circu}).  To the lowest order these solutions reduce to
\begin{equation}
\eta \simeq A_1 \tau, \ \ \ \delta x \simeq - {C
\over A_1} {\rm cos} A_1 \tau.
\end{equation}
Near the secular resonance where $\vert A_1 \vert$ is relatively small,
the amplitude of $\delta x$ becomes large.

2) {\it Libration.} In the opposite limit that $\vert A_1 \vert \ll
\vert A_2 \vert$, the non axisymmetric secular interaction between the
planets is important.  There are stationary points in the ($x$-$\eta$)
plane which center on the values of $x=x_m$ and $\eta =0$ or $\pi$
(see Appendix \ref{sec:libra1}). Around these points, there are orbits with
both small amplitude modulation such that
\begin{equation}
\epsilon \equiv {x - x_m \over x_m} = \epsilon_0 {\rm sin} \omega \tau,
\ \ \ \ \ \eta = \eta_0 {\rm cos} \omega \tau,
\label{eq:epsil2}
\end{equation}
where
\begin{equation}
\eta_0 = \pm \epsilon_0 \ \ \ \ \ {\rm and} \ \ \ \ \ \
\omega = \pm {C  \over x_m}
\left( 1+ \gamma_d x_m^2  \right)
\label{eq:ome}
\end{equation}
is the oscillation frequency (see Appendix \ref{sec:libra1}).  The
dimensionless amplitude of these librational orbits $\epsilon_0 \ll 1$.
For these orbits, although $g_1 \ne g_2$, secular interaction induces
them to precess at similar frequencies that their relative longitude
of periapse passage is always approximately aligned or anti aligned.
Thus, these planets are effectively in a state of secular resonance.
Note that because the libration is centered around $\eta = 0$ or $\pi$,
there is no effective angular momentum transfer despite the phase lock.

3) {\it Excitation.}  For systems with $\vert A_1 \vert < \vert A_2
\vert$, there are also orbits in which $x$ and $\eta$ have very
different values as $x_m$ and $0$ (or $\pi$) respectively.  In these
cases, $\eta$ would evolve rapidly to a phase angle
\begin{equation}
\eta_1 \simeq {\rm cos}^{-1} \left(- {A_1 \over A_2} \right)
= {\rm cos}^{-1} \left( {-A_1 x \over C (1 - \gamma_d x^2) } \right),
\label{eq:eta1}
\end{equation}
such that $d \eta / d \tau $ is reduced to zero (see eq. [\ref{eq:detadt}])
and the two planets become phase locked.

For all non zero (or $\pi$) values of $\eta_1$ , the monotonic
increases/decreases of $x$ correspond to eccentricity
excitation/damping, analogous to the situation of precessional
degeneracy.  In this state of near secular resonances, angular
momentum is monotonically transferred from one planet to the other,
resulting in a monotonic evolution of $x$. The modification of $x$ in
turn leads to an evolution in $\eta$.  Along the path of $x$ and
$\eta_1$ evolution, $g_1 \ne g_2$ and the planets would librate about
their evolving guiding center. For the special cases when the two
planets enter the resonance, the second
order solution reduces to that in equation (\ref{eq:epsil2}) and all orbits
become instantaneously librational, even though $\epsilon_0$ and
$\eta_0$ may become arbitrarily large.  At the center of resonance
where $A_1$, $A_2$, and $A_1/A_2$ all vanish, $\eta_1=\pi/2$ and $x_m$
evolves exponentially.  We discuss the evolution of these systems in
Appendix \ref{sec:deplet}.

\section{MODELS}
The above analytic approximation is useful for isolating the three
family of orbits along with two stationary points in the ($x$-$\eta$)
diagram.  These orbits generally follows the contours of equi
Hamiltonian map (see Appendix \ref{sec:hamil}).  However, the
topological evolution of the Hamiltonian map alone is insufficient for
the determination of the planets' orbits during the depletion of the
disk.  We carry out, below, numerical integration of the {\it full}
equation of motion for the planets subject to the potential of their
host star and nascent disks.  In this section, we briefly describe a
model prescription with which we examine the passage of librational
degeneracy during the epoch of disk depletion.

\subsection{Planetary Formation Scenarios and Disk Model}

We first discuss the physical process of disk depletion.  According to
conventional theories, planets are formed through the condensation of
grains which grow to planetesimals via cohesive collisions (Hayashi,
Nakazawa, \& Nakagawa 1985; Lissauer 1987; Wetherill 1990).  Upon
attaining a sufficiently large mass, planetesimals accrete gas (Mizuno
1980; Bodenheimer \& Pollack 1986; Pollack et al. 1996).  Eventually,
their growth is terminated when protoplanets can tidally induce the
formation of a gap near their orbit (Goldreich \& Tremaine 1980; Lin
\& Papaloizou 1980, 1993; Takeuchi, Miyama, \& Lin 1996).

Thereafter, gas in the inner of the disk continues to diffuse inward
as it loses angular momentum to the planet. Since gas replenishment is
cut off by the formation of the gap, the inner region is depleted well
before the outer region of the disk. For the present discussions, we
assume that the mass in the inner regions of the disk becomes
negligible by the stage when the second planet is fully grown. Thus,
in our analysis of planets' dynamical evolution, we only need to
consider their interaction with the disk region extended outside the
outer planet. Both the gravitational and tidal influences of the disk
on the outer planet are much more intense than on the inner planet
because the former is much closer to where most of the gas is
distributed.

When the planets' orbits lie in the same plane as their nascent disks
and their distance to the nearest edge of their nascent disks is
greater than the disks' scale height, a thin-disk approximation is
sufficient for the computation of the disks' gravitational potential.
Following the approach of Ward (1981), the self-gravitating potential
for a disk model with a surface density profile $\Sigma= \Sigma_0
(r_0/r)^k$ (with $\Sigma_0$ and $r_0$ being some fiducial values) can
be expressed as
\begin{equation}
V(r)=2\pi G \Sigma_0 r
\left(\frac{r_0}{r}\right)^{k}\sum_{n=0}^{\infty}
\left(\frac{A_n}{2n+k-1}\right)\left( \frac{r}{r_{\rm
edge}}\right)^{2n+k-1}. \label{eq:vpot}
\end{equation}
In the above expression, $r$ is the instantaneous location (rather
than the semi major axis) of the planet.  We also assume that the disk
has an inner edge which is located at $r_{\rm edge}$, interior to
which $\Sigma=0$.  The coefficient $A_n=\{(2n)!/2^{2n}(n!)^2\}^2$. For
illustration purpose, we consider the minimum mass solar nebula model
(Hayashi 1981), in which $k=3/2$.  The general applicability of this
phenomological model for protostellar disks around different stars is
questionable.  In the absence of a reliable prescription, we
parameterize disk models with a range of values in $k$ (see \S4.3).  We
assume the disk is extended to infinite and the semi-major axes of
planets do not change as a consequence of disk depletion or planets'
secular interaction with each other. The results to be presented below
show that while the overall dynamical evolution of the planets' orbit
is insensitive to the detailed disk model, the planets' extrapolated
initial eccentricity ratio does depend on the mass of the disk gas
near their orbits.

\subsection{Planetary System and Prescription for Disk Depletion}

For computational simplicity, we consider a planetary system consists
of two planets with coplanar orbits.  In the case of the Upsilon
Andromeda system, we are concerned with gravitational interaction
between planets c and d and neglect the contribution by planet b.  We
also neglect the influence of the planet's tidal interaction with the
disk. Only the axisymmetric component of the disk potential is taken
into account.  Mutual inclinations between the two planets and between
the planet and the disk are neglected.  The inclination of the orbital
plane is denoted by $i$.  We take surface density of the disk as
$5\times \sin i$ times that of the minimum mass solar nebula model.
Only the ratios, rather than the individual values, of the planets'
and the disks' masses determine the planets' motion.  Thus, our
results do not depend on the inclination of the system (except for
small changes in the mean motion and the long term dynamical stability
for exceptional small values of $i$).

For models with different values of $k$, we scale $\Sigma$ such that
its value at the disks' inner edge $r_{\rm edge}$ remains invariant.
As we have indicated above, inner region of the disk is likely to be
severely depleted prior to the emergence of the second planet.  Thus,
we assume the disk depletion to proceed in an inside out manner by
adopting a model in which the location of nebula inner edge ($r_{\rm
edge}$) is prescribed to expand outwards as $r_{\rm edge}(t)=r_{\rm
edge} (t=0) + r_i t/t_{\Delta N}$ with $r_i = 1 {\rm AU}$ while
$\Sigma_0$ is kept constant. In order to demonstrate that the passage
of the secular resonance does not depend on the detailed prescription
of the depletion process or the structure of the disk, we also
consider a model in which the surface density of the disk is assumed
to decline by an identical reduction factor everywhere such that
$\Sigma_0(t) =\Sigma_0(t=0) \exp(-t/t_{\Delta N})$ with $r_{\rm
edge}=$constant.  In the limit that $a_c/a_d \ll 1$ and
$a_{c,d}/r_{\rm edge} \ll 1$, we find that the two prescriptions of
the disk depletion do not lead to a significant difference in the
numerical results because $\Delta N \propto \Sigma/r_{\rm
edge}^{5/2}$.  For the approximate prescription of $\Delta N$ (see
eq. [\ref{eq:dnexpsit}]), the magnitude of $A_1 =1+\Delta N -\gamma_d$
changes on a time scale $\tau_{A_1} \sim t_{\Delta N}$ and the
condition of the secular resonance $A_1 =0$ is satisfied when
\begin{equation}
\frac{\Sigma_0 r_0 ^{3/2} }{r_{\rm
edge}^{5/2}}\simeq
\frac{5(\gamma_d-1)}{4\pi}\frac{M_d}{a_d^3}\left(1-\frac{n_c}{n_d
}\right)^{-1}.
\label{eq:condsreso}
\end{equation}
This condition can be attained with both prescriptions. 

The orbital evolution of the planets can be directly integrated
numerically, including the point-mass potential of the host stars and
sibling planets as well as the contribution from the disk potential
where the gravity can be expressed as
\begin{equation}
F=\frac{dV}{dr}= 2\pi G \Sigma(r,t) \sum_{n=0} \frac{4n }{4n+1} A_n
\left(\frac{r}{r_{\rm edge}(t)}\right)^{2n+1/2},
\end{equation}
where $k=3/2$ is assumed.

\section{RESULT OF NUMERICAL COMPUTATION}

\subsection{Evolution of Librating Planets Around Ups And}
Using the equi-Hamiltonian contour maps, we illustrate the orbital
evolution of planets c and d during disk depletion.  Figure
\ref{fig:upshmp} shows the time evolution of the equi-Hamiltonian map
of Upsilon Andromeda system.  In these calculations, we adopt the mass
ratio $M_{c}/M_{d}=0.502$ based on the assumption that they occupy the
same orbital plane.  The nebula edge retreats from inside to outside
(from panel 1 to panel 5).  When the nebula edge is at 4.2AU (panel
1), the orbits librate around ($x=10$, $ \eta=0$) or ($x=0.4$, $
\eta=\pm \pi $).  The precession speed of the longitude of planet c's
periastron is slower than that of planet d's periastron except for
the occasional regressions.  Embedded within such a disk, a pair of
planets with the present observed values of $x$ and $\eta$ would not
follow a closed librating track.  When the nebula edge is at 5AU
(panel 2), the two eigenfrequencies of the system become closer to
each other.  The eccentricities of planets modulate with large
amplitudes.  The magnitude of $x$ librates on a closed
equi-Hamiltonian track, about a central point at $x \sim 3.3$ and
$\eta=0$.

From the estimation of the equation (\ref{eq:dnexpsit}), we find that
$A_1 = 1+\Delta N-\gamma_d$ vanishes when the disk is retreated to
$r_{\rm edge}=5.5$AU, namely, secular resonance occurs at $r_{\rm
edge}=5.5$AU.  The panel (3) represents the epoch of secular resonance
passage.  At this stage, the precession frequencies of the longitudes
of periastrons nearly match with each other ({\it i.e.},
$\dot{\eta}\sim 0$ compared with $\dot{x}$).  Consequently, $\eta$
varies slowly (with finite values other than 0 or $\pi$) in comparison
with the non negligible evolution of $x$ such that the eccentricities
can change significantly.  The equi-Hamiltonian map is locally
symmetric about the lines of $\eta= -\pi$, $-\pi/2$, $0$, $\pi/2$, and
$\pi$ and that of $x=\gamma_d^{-1/2} \sim$2. 

When the nebula edge is beyond 5.5AU, the precession speed of
longitude of planet c's periastron is faster than that of planet d's
periastron (except for the occasional regressions). Panel (4) shows
the case that the nebula edge has retreated to 7AU. The centers of the
closed libration tracks are reversed from the situation in panels (1)
and (2). The planetary configuration with $e_{c}>e_{d}$ and $
\varpi_{c}\sim \varpi_{d} $ in panel (2) becomes that with
$e_{c}<e_{d}$ and $ \varpi_{\rm c}\sim \varpi_{\rm d} $ in panel
(4). When the entire protoplanetary disk is depleted (panel (5)),
stable orbits librate around ($x=0.5$, $\eta=0 $) or ($x=7$, $\eta=\pm
\pi$). The shape of equi-Hamiltonian map is turned upside down with
$x=\gamma_d^{-1/2}$.  Figure \ref{fig:hmpupsuni} is the time evolution
of the equi-Hamiltonian map for the uniform-depletion prescription
with $r_{\rm edge}=4.5$AU.  In this model, the secular resonance
occurs when $\Sigma_0(t)/\Sigma_0(t=0) =0.5$.  Although the time of
the secular-resonance passage is slightly modified from that obtained
with the model of inside-out depletion, the general evolutionary
pattern is not qualitatively changed.

We also numerically integrate the orbits of planets c and d using the
the same inside-out disk depletion prescription as above ({\it i.e.},
we arbitrarily set the initial disk edge to be at 4.2 AU and specify
its retreating speed to be $10^{-5}$AU ${\rm year}^{-1}$).  We consider two sets of
initial conditions.  Figures \ref{fig:upseome}a and c illustrate the
evolution of a model with initial values of $x=8$ and $\eta=\pi/4$.
We illustrate the orbital evolution of planets c and d with the
($x$-$\eta$) diagram (see Fig. \ref{fig:upseome}a) and with $e_c$ and
$e_d$ as a function of time (see Fig. \ref{fig:upseome}c).  These
results show that the planets' orbits initially librate relative to
each other.  The planets remain on librating closed tracks in the
($x$-$\eta$) diagram after the disk is totally depleted.  But, as a
consequence of the secular-resonance passage, the eccentricity of
planet c becomes smaller than that of the planet d.  The time-averaged
eccentricities after the disk depletion are $\langle e_{c}\rangle
\simeq 0.12$ and $ \langle e_{d}\rangle\simeq0.2$.

We also consider a second set of initial conditions with $x=8$ ($e_{c}
=0.4$, $e_{d}=0.05$) and $\eta =3\pi/2$. In this case, the planets'
orbits initially circulate relative to each other, but, they become
trapped in librating orbits during the passage through the secular
resonance (see Fig. \ref{fig:upseome}b).  The mean eccentricities
after the total depletion of the disk are very similar to that
obtained in the case of panel (a), because they are determined
by the conservation of angular momentum (eq. [\ref{eq:angcon}]).  

We also carried out numerical integration of the full equations with
the uniform-disk-depletion prescription with $t_{\Delta N}= 10^5$
years, $r_{\rm edge}=4.5$AU. When the same initial condition is used
({\it i.e.} $x=8$ and $\eta=\pi/4$) were used, the pattern of the
orbital evolution with this prescription (see Fig. 
\ref{fig:upseomeuni}a) is not significantly changed from that in
Figure \ref{fig:upseome}a.  In panels b and c, we show the evolution
from ($x=8, \eta=\pi$) and ($x=5.6, \eta=0$).  The pair of planets in
panel b has an initially wide open circulating track.  During the disk
depletion, they temporarily enter into closed librating track.  But,
their orbital configuration becomes wide open again when most of the
disk material is depleted.  The results of these numerical
calculations show that as long as the depletion time scale of the disk
is longer than the oscillation period of eccentricity, the planets'
orbits evolve adiabatically.  Those systems with librating orbits in
($x$-$\eta$) diagram prior to the disk depletion usually remain on the
closed librating tracks after disk is totally depleted despite large
changes in magnitude of $x$ as in the case of Ups And (see below).
Those pairs of planets with the widely open circulating tracks
initially generally remain on open circulating tracks after the disk
depletion is completed as in the case of HD168443 (see below).  Under
some circumstances, it is also possible for planets with marginally
circulating/librating initial orbits to undergo transition after the
disk depletion.

\subsection{Orbital Evolution of Circulating Planets Around HD168443}
Next we consider the case of the planetary system around HD168443 for
which we assume the mass ratio to be $M_{b}/M_{c}=0.451$.  Figure
\ref{fig:hdhmp} shows the time evolution of the equi-Hamiltonian
contour map of the HD168443 system.  The nebula edge retreats from
inside to outside (from panel 1 to panel 5).  The secular resonance
occurs when $\Delta N=-0.86$ with $r_{\rm edge}=7.2$AU.  Similar to
the case of the planetary system around Upsilon Andromeda, the
librating orbit with $\vert \eta \vert<\pi/2$ are confined in regions
with $e_{b}>e_{c}$ prior to disk depletion.  But, in the case of
HD168443 system, the domain of closed librating tracks is small.  (In
contrast, for small $ M_{b}/M_{c}(a_{b}/a_{c})^{1/2}$, the domain of
open circulating tracks is large as is the case for planets c and d
around Ups And).  The domain of closed librating tracks with $\vert
\eta \vert<\pi/2$ moves downward in the Hamiltonian map as disk
depletion proceeds.  If planets of HD168443 initially follow open
circulating tracks prior to the disk depletion, they would remain on
the open circulating tracks after the disk depletion.  In this case,
during the onset of secular resonance, planets b and c may briefly
attain a librating track before moving onto an open circulating track
as the disk continues to deplete.  Throughout the epoch of disk
depletion, the planets' longitudes of periastrons are widely separated
such that little angular momentum may be exchanged between them.
Consequently, the net change of eccentricities is limited.  

Figures \ref{fig:hduni} show the numerically computed orbital
evolution.  The disk potential is calculated with the uniform
depletion prescription in which we specify $t_{\Delta N}=10^6$ years.
Figures \ref{fig:hduni}a and b represent the first model with $x=6$
and $\eta=0$ initially and Figures \ref{fig:hduni}c and d show the
second model with $x=1$ and $\eta=0$ initially.  For the first model,
the eccentricity of outer planet c becomes larger than that of planet
b after the disk depletion in contrast to the observed eccentricities
of planets b and c.  With the initial $x$ smaller than
$\gamma_d^{-1/2}$, the results for the second model (in panel c)
reproduce the present eccentricities of both planets.  (For
presentation simplicity, the notation of $\gamma_d = (a_b/a_c)^{1/2}
(M_b/M_c)$ which represents planets b and c.)  The extent of the open
circulating track is large both before and after the disk depletion.
During the passage through the secular resonance, the planets' orbits
are temporarily trapped in the libration cycles around $\eta=\pi$.
But the two planets break free from their librational state to follow
a circulation path after the disk is totally depleted.

\subsection{Dependence on Disk Model}

\subsubsection{Single External Disks}
The initial eccentricities of planets, prior to the disk depletion,
can be inferred from the conservation of the total angular momentum
and energy of the planets.  We represent the mean eccentricities
($<e_c>$ and $<e_d>$) of the planets c and d around Ups And by their
values near the libration center ({\it i.e.}  with $x=x_m$ and $\eta
=0$).  In Figure \ref{fig:modeldep}a, we plot the values of $<e_c>$
and $<e_d>$ for various $\Sigma_0$ and $r_{\rm edge}$.  The
eccentricities $<e_c>$ and $<e_d>$ are shown by filled circle for six
different pre-depletion disk surface densities ($5\sin i$, $4\sin i$,
$3\sin i$, $2\sin i$, $1\sin i$, and zero times that of the minimum
mass model at $r_{\rm edge}$ where zero corresponds to the present
state).  The model with $k=3/2$ has the same power-law index for the
$\Sigma$ distribution as that of the minimum mass nebula model so that
these scaling factors for $\Sigma (r_{\rm edge})$ also represent that for
the disk mass.  

We also adopt three different pre-depletion locations of the edge
($\log(r_{\rm edge}/a_d) =0.14$, $0.2$, $0.3$). The pre-depletion
locations of the edge correspond to 3.53, 4.06, and 5.11 AU,
respectively.  The apogee distance of planet d ($a_d (1 + e_d)$) is
unlikely to exceed the inner edge of the outer disk ($r_{\rm edge}$)
because its eccentricity may be effectively damped and the disk may be
strongly perturbed (Papaloizou et al. 2001).  From this
requirement, we note that $r_{\rm edge} >3.5$AU for the present value
of $e^{\prime}_{d} = 0.35$.  Prior to the disk depletion, it is
possible that $e^{\prime}_{d} $ may be close to zero.  Nevertheless,
the tidal torque of planet d can induce the formation of a gap with a
width which is $\Delta r \sim {\sqrt 12} R_{\rm Roche}$ where $R_{\rm
Roche} = (M_d/ 3 M_\ast) ^{1/3} a_d \simeq 0.35 a_d$ is its Roche
radius (Bryden et al. 1999).  Thus, we adopt the minimum value
of $r_{\rm edge} \simeq a_d + \Delta r \simeq 3.5$AU (see top panel of
Fig. \ref{fig:modeldep}a).

The solid line shows an ellipse which is delineared by the
conservation of total angular momentum (see eq. [\ref{eq:angcon}] in
Appendix).  To second order in eccentricity, the conservation of the
total angular momentum (eq. [\ref{eq:ang}] in Appendix) implies that
\begin{equation}
\frac{e^{\prime 2}_{d}- e_{d}^2}{ e^2_{c}- e_{c}^{\prime 2}}
= \frac{M_{c}\sin i \sqrt{a_{c}}}{M_{d}\sin i \sqrt{a_{
d}}}= \gamma _d,
\label{eq:daen}
\end{equation}
where $e^{\prime}_{c}$ and $e^{\prime}_{d}$ are the pre-depletion
eccentricities of planets c and d respectively.  The value of $\gamma
_d \sim 0.286$ in the case of planets around Ups And. A maximum
value of $e^{\prime}_{d} = 0.37$ is attained for a minimum value of 
$e^{\prime}_{c} = 0$. But, our results in Figure \ref{fig:upseome} indicate
that $x$ decreases during the disk depletion so that $e^{\prime}_{c}
> e_c$ and the maximum value of $e^{\prime}_{d}$ is unattainable.

When the disk edge is relatively close to the planet (top panel),
$e^{\prime}_{c} >0.65$ and $e^{\prime}_{d} < 0.1$ if the pre-depletion
disk mass is greater than $2\sin i$ times that of the minimum mass
nebula model.  But for larger values of $r_{\rm edge}$, the
pre-depletion eccentricities approach their present values as the
contribution of the disk mass to the total gravity becomes vanishingly
small.

The secular resonance occurs at $x_m \sim2 $ in the case of Ups And.
As long as there was the disk which corresponds to a pre-depletion
$e_c>0.51$, the planets pass through the secular resonance.  Note that
with an arbitrarily large disk mass, $e^{\prime}_{d}$ vanishes while
$e^{\prime}_{c}$ attains a maximum value of $0.69$.  Even under these
extreme condition, the orbits of the two planets do not cross and
remain stable. During the disk depletion, the eccentricities of planet
c and d coincide with each other at about 0.33.  For the pre-depletion
eccentricity of planet c to be $>0.33$ (which is generally the case
with the exception of the three least massive initial disk models),
the inversion of eccentricity ratio between the planet c and d around
Ups And occurs.

Because the influence of the disk on the planet is mainly due to the
gas near the $r_{\rm disk}$, the above results are insensitive to
changes to the surface density distribution at large radii.  In Figure
\ref{fig:modeldep}b, we set $r_{\rm edge}= 4.057$AU ($\log(r_{\rm
edge}/a_d) =0.2$) and considered three sets of $\Sigma$ distributions
with $k=2$, $k=1$, and $k=1/2$ (Fig. \ref{fig:modeldep}b).  We adopt
six sets of values of the surface density $\Sigma (r_{\rm edge})$ of
each model at $r_{\rm edge}$.  For disk with similar surface density
scaling factor, there are no significant differences between three
panels.  The detailed disk structure is not important for the
evolution of the planetary eccentricities in our model.

\subsubsection{Interplanetary Rings and Outer Disks}
After the emergence of multiple planets, gap formation effectively
cuts off gas flow from the external disk across the outermost planet's
orbit.  Since both dynamical and viscous time scales are rapidly
increasing function of the disk radius, the disk interior to the gap
is depleted rapidly (Lin \& Papaloizou 1986b).  Due to a shepherding
effect induced by the multiple planets (Goldreich \& Tremaine 1979),
the gaseous interplanetary rings may be preserved for a much longer
time scale than the internal disk.  Nevertheless, the dissipation of
the competing tidal perturbations by the planets can lead to ring
dispersal well before the depletion of the external disk (Bryden 
et al. 1999).  

In the present context, the evolution of the eccentricity ratio
($x_{m0}$) at the libration center around $\eta=0$ depends intricately
on the depletion pattern of both the ring and disk as well as the
locations of the edges of the ring and disk.  For example, the value
of $x_{m0}$ is smaller for interplanetary rings with inner edge
$d_{\rm in}$ closer to the inner planet.  In contrast, the value of
$x_{m0}$ is larger for interplanetary rings with outer edge $d_{\rm
out}$ closer to the outer planet.  The balance between these effects
determines the evolution of the eccentricity ratio.  In order to
illustrate these dependence, we consider the epoch when planets c and
d around Ups And were surrounded by an interplanetary ring and an
external disk.  We adopt a surface density distribution which is
$5\times \sin i$ that of the minimum mass solar model.  A hole is cut
out interior to $d_{\rm in}$.  An annular strip is also removed
between $d_{\rm out}$ and $r_{\rm edge}$.  By fixing the inner edge of
the external disk to be $r_{\rm edge} =4$ AU, we compute the value of
$x_{m0}$ as a function of $d_{\rm in}$ (see Fig. \ref{fig:ring}). In
these calculations, we used eight values of the outer edge of the ring
$d_{\rm out}$, ranging from 1.2 to 1.9 AU in intervals of 0.1AU.

The results in Figure \ref{fig:ring} indicate that when $d_{\rm out}
<$ 1.66AU, the gravitational perturbation of the ring on the inner
planet always surpasses that on the outer planet.  Consequently, the
value of $x_{m0}$ is reduced below that ($x_{\rm disk}\sim 12$) in the
external-disk-only ({\it i.e.} ringless) models.  In these small
$d_{\rm out}$ cases, $x_{m0}$ approaches to 12 for relatively large
$d_{\rm in}$ as the ring's influence weakens. But, in the limit
$d_{\rm out}>$1.66AU, $x_{m0}$ becomes larger than $x_{\rm disk}$ for
relatively large $d_{\rm in}$.  For example, when $d_{\rm out}=$1.8AU,
$x_{m0}$ becomes larger than $x_{\rm disk}$ at $d_{\rm in}>$1.5AU
because the ring's perturbation on the outer planet is enhanced.  Also
in this large $d_{\rm out}$ limit, $x_{m0}$ becomes smaller than
current observed eccentricity ratio of 0.66 at $d_{\rm in}<$1.33AU.
For all values of $a_{\rm out}$, the pre-depletion values of $x_{m0}$
in this model is always larger than the current observed eccentricity
ratio.

In Figure \ref{fig:ringhmp}, we show the evolution of equi-Hamiltonian
contours in a model with both ring between planets c and d and an
external disk beyond planet d.  We adopt a prescription in which the
ring is depleted uniformly prior to the uniform depletion of the
external disk.  For our model, we specify $\log(d_{\rm in}/a_c) =
\log(a_d/d_{\rm out})$=0.2 for the ring and $\log(r_{\rm edge}/a_d)$
=0.2 for the external disk.  Our numerical results indicate that
$x_{m0}$ is initially smaller than current observed eccentricity ratio
(panel 1).  The ring proceeds to be depleted during the epochs
represented by panels (2) and (3). When the ring mass is reduced to
half of its original value, $x_{m0}$ become $\sim 2$ (panel (2)), {\it
i.e.} the system passes through a state of secular resonance.  The
entire ring is depleted while the external disk is preserved during
the epoch represented by panel (3).  After that, the contours again
evolve through a state of secular resonance (when 1/3 of the external
disk is depleted in panel 4) to current values (in panel 5) as in the
case of Figure \ref{fig:hmpupsuni}.  

Although the same pattern of evolution occurs in the case of planets
around HD168443, there is little change in the eccentricities of
either planets since the contour of their eccentricity ratio always
retains the values around $\gamma_d^{-1/2}$.  We realize that the
evolution of the interplanetary ring is determined by the process of
gap formation and planet-disk tidal interaction, its influence on the
eccentricity evolution of the planets should be analyzed along with
its non axisymmetric contributions.  The results of such analyses will
be presented elsewhere.

\section{IMPLICATIONS ON THE PROTOPLANETS' ORBITAL ECCENTRICITY PRIOR 
TO THE DEPLETION OF THEIR NASCENT DISKS}

\subsection{Planets around Ups And}
First, we consider the evolution of planets c and d around Ups And.
We infer from their present librating orbits and the results in Figure
\ref{fig:upshmp} that these planets followed a close librating track
prior to the disk depletion.  Our results also indicate that when the
disk dominated the planets' apsidal precession, the domain of close
librating tracks with $\vert \eta \vert <\pi/2$ resided in the
parameter region with $e_{c}>e_{d}$.  Thus, the present observed
librating orbits of planets c and d would be attainable if $ e_{c}>
e_{d}$ prior to the disk depletion.  Since $ e_{c}< e_{d}$ today, the
passage through the secular resonance must have led to the inversion
of this ratio.  In Figure \ref{fig:upsang}, we map out, in the
($x$-$\eta$) plane, the most likely initial domain which may have led
to the present planetary orbits in the Ups And system (panel a).  The
periapses of those pairs of planets which initially occupied region 1
maintain their near alignment despite the large changes in $x$ during
the disk depletion.  These systems enter the domain of closed
librating tracks in the ($x$-$\eta$) plane after the disk is
completely depleted.  Their orbits become very similar to the present
orbits of planets c and d.  Planets in region 2 have open circulating
tracks prior to the disk depletion.  Their orbits remain on
circulating tracks after the disk depletion. Among those pairs of
planets which originated from both regions 1 and 2 ($x>\gamma_d^{-1/2}
\sim 2$), the eccentricity ratio is reduced after the disk depletion.
Thus, we infer those pairs of planets which currently occupy the
region $x (=e_c/e_d) \lesssim 1$ are originated from the domain $x
\gtrsim 2.5$ prior to the disk depletion.  Those pairs of planets with
$x<2$ initially cannot attain the present observed values of $x (\sim
0.6)$ because the magnitude of their $x$ increases during the disk
depletion.

In the absence of a strong non axisymmetric torque associated with
the planet-disk interaction, secular interaction between the planets
does not alter the total angular momentum and energy of the system.
In contrast to the eccentricity ratio ($x$) between planets c and d,
the magnitude of eccentricities cannot be excited by the sweeping
secular resonances alone. The equation (\ref{eq:daen}) indicates that
if both planets had nearly circular orbits initially, their
eccentricity would remain small after the disk is completely depleted.
Thus, around Ups And, at least planet c must have acquired some
initial eccentricity prior to the epoch of planet formation.

Up to now, we have neglected the effect of non axisymmetric torque
associated with the planet-disk interaction.  During their formation,
the eccentricities of isolated protoplanets are damped and excited as
a result of their interaction with their nascent disks at the
corotation and Lindblad resonances respectively.  For protoplanets
with masses comparable to that of Jupiter, the damping effect of the
corotation resonances is stronger than the excitation effect of the
Lindblad resonances (Goldreich \& Tremaine 1980).  But protoplanets
with masses an order of magnitude larger than that of Jupiter may open
relatively wide gaps in protostellar disks.  In this limit, the
protoplanets' corotation resonances may be cleared of disk gas such
that their eccentricity may be excited (Artymowicz 1993; Papaloizou,
et al. 2001).  For modest inclination angle between the
their orbital plane and the line of sight, neither planet c nor d have
sufficiently large mass to clear a wide gap.  But, planet c is likely
to form prior to planet d.  The tidal torque of planet d may truncate
the disk well beyond the region which contains any corotation
resonance for planet c so that the latter's eccentricity is excited
whereas that of planet d is damped.

The actual mass of planets c and d depends on the value of sin $i$.
If $i < 20^{\circ}$, the mass of planet d would be sufficiently large
for its interaction with the disk to excite eccentricity.  The
constraint on the inclination angle inferred from the dynamical
stability of the planetary system around Ups And indicate that $i >
12-45^{\circ}$ (Lissauer \& Rivera 2001; Ito \& Miyama 2001). In a
multiple system, however, the rate of angular momentum transfer
between planet d and the outer disk may be faster than that between it
and planet c.  We shall consider in a future contribution whether
planet c will be able to attain a larger initial eccentricity.

It is also possible that the planets around Ups And was closer to each
other in the past and dynamical instability may have led them to
undergo orbit crossing (e.g., Chambers, Wetherill, \& Boss 1996).
Rasio et al. (1996) suggested that such a process may have led to the
scattering of a planet close to the surface of its host star and
subsequent tidal circularization may have caused it to become a short
period planet.  Around Ups And, regardless whether such a process can
lead to the formation and survival of the short-period planet b,
planetary scattering may also have excited the eccentricity and
modified the semi-major axes of both planets c and d (
Weidenschilling \& Marzari 1996, Lin \& Ida 1997).  During such encounters,
planets with lower mass generally acquires larger eccentricity as a
result of the conservation of the system's total energy and angular
momentum.  Although these speculations provide arguments for planet c
to have initially acquired more eccentricity than planet d, rigorous
analysis is needed to quantitative verify these conjectures.

\subsection{Planets around HD168443}

In the case of planets around HD168443, the present orbit follows an
open circulating track.  In Figure \ref{fig:upsang}b, we outline the
necessary initial domain in the ($x$-$\eta$) diagram, which would lead
planets b and c to acquire their present orbital parameters.  Planets
which were initially in the region 2, would enter into open
circulating tracks after their nascent disk is completely
depleted. But their eccentricity ratio would not generally match those
inferred from the observations of HD168443.  The most likely initial
domain of planets around HD168443 (with their current dynamical
properties) is mapped out as region 1. During the passage through
secular resonance, the eccentricity of planet c decreases while that
of planet b increases slightly.  The geometry of ($x$-$\eta$) diagram
is reversed during the passage through the secular resonances, {\it
i.e.}, at the point $x=\gamma^{-1/2}_d$.  This small change is due to
the present value of $x$ being close to that of $\gamma^{-1/2}_d$ in
the HD168443 system.  In contrast, present $x$ of planets around
Upsilon Andromeda is much smaller than $\gamma^{-1/2}_d$.

\subsection{Other Multiple-planet Systems}

We comment here about another system of planets around HD74156 (Table
\ref{tab:parameta}(c)). In this system, the outer planet is more
massive than the inner planet.  Since their orbits are circulating
(similar to those of the planets around HD168443), the evolution of
this system during the disk depletion would be very similar to that
described in \S4.2 for the planets around HD168443. However, the
present value of $x$ is not close to that of $\gamma^{-1/2}_d$.  The
inversion of the eccentricity ratio is expected to occur in the case
of planets around HD74156 (similar to those of the planets around Ups
And).

Our own solar system also contains multiple gaseous giant
planets. In this case, the inner planet, Jupiter, is more massive than
the outer planet, Saturn.  Their longitudes of periastron are not
aligned with each other.  The ratio of the Jovian eccentricity to the
Saturnian eccentricity ($e_{\rm J}/e_{\rm S}$) varies between the
range of 0.3 and 5 with time.  In this case, the orbits of Jupiter and
Saturn do not pass through the secular resonance during the depletion
of the primordial solar nebula.  Therefore, the initial orbits
attained by Jupiter and Saturn during their formation would not be
significantly changed as a consequence of the disk depletion.  This
evolution pattern is similar to that of the two-planet system around
47 UMa.  Those two planets have masses 2.0 and 0.67 ${M_J}$ and
semi-major axes 2.1 AU and 3.8AU, respectively (Table
\ref{tab:parameta}(d)).

In general, those systems where the outer planets are more massive
than the inner planets (such as the planetary systems around Ups And,
HD168443, and HD74156), the depletion of the outer disks induces a
secular resonance which sweeps through both planets.  But in the
opposite situation where the inner planets are more massive (more
precisely, where $\gamma_d^{-1/2}<1$) such as the planetary systems
around the Sun and 47 UMa, the depletion of the disk does not lead to
the secular resonances between the planets.  Thus, the changes of the
eccentricities are not so large.  The planets around HD12661 (Table
\ref{tab:parameta}(e)) have librating orbits with $\eta \sim \pi$.
During the disk depletion, the secular resonance passes through this
system ({\it i.e.} $x_m$ attains a value of $\gamma_d^{-1/2}$) because
its $\gamma_d^{-1/2}>1$ (the inner planet is massive than the outer
planet). Today, the value of $x_m$ around the libration centers at
$\pi$ is slightly larger than $\gamma_d^{-1/2}$.  Thus, the
pre-depletion value of $x_m$ must be smaller than $\gamma_d^{-1/2}$.
Nevertheless, the magnitude of the eccentricity change is relatively
small because the present value of $x_m$ is close to that of
$\gamma_d^{-1/2}$.

\section{SUMMARY AND DISCUSSIONS}

In this paper, we consider the secular interaction between pairs of
coplanar planets.  Provided these planets are not locked in a mean
motion resonance, their secular interaction leads to angular momentum
exchange without energy transfer between them.  Consequently, both
planets undergo apsidal precession and eccentricity modulation about
some mean values.  Thus, dynamical evolution of these systems of
planets is best illustrated by the ratio of their eccentricity $x$ and
the relative longitude of their periastrons $\eta$.  For example,
based on their observed orbital elements, we find that gravitational
perturbation between the outer two planets around Ups And causes both
$\eta$ and $x$ to modulate, with small amplitude, along a close
librating track around $\eta=0$ and $x\sim 0.66$ (see Fig.
\ref{fig:orbitnow}).

Although the mean eccentricities of these planets are constant today,
they may have undergone significant changes during the epoch when
their nascent disks were depleted.  In order to extrapolate the
initial dynamical properties of planetary systems, we consider the
precession caused by the self gravity of these disks.  We show that
contribution from the disks initially dominates the precession
frequency of the planets and regulates the angular momentum exchange
rate between them.  As their influence fades during the disks'
depletion, the precession eigenfrequencies of the planets' longitude
of periastron pass through a state of secular resonance.

As a result of this transition, the magnitude of both planets'
eccentricities changes with their ratio inverted.  But, provided the
disk depletion proceeds on a time scale long compared with the
precession frequencies associated with the secular interaction between
the planets, the libration amplitude of $\eta$ is not significantly
changed due to the preservation of an adiabatic invariance.  Thus, if
a pair of planets started out with a close librating orbit prior to
the disk depletion, they would generally retain the librating nature
of their secular interaction after the disk depletion.  Planets with
open circulating tracks initially would generally evolve to attain
other open circulating tracks after the disk depletion.  However, the
circulating track may temporarily enter into closed librating track
near the secular resonance.  Planets with marginally circulating
orbits (close to librating orbits) prior to the disk depletion may
sometimes attain librating orbits after the disk depletion provided
the difference between the libration and resonant centers ({\it i.e.}
$ \vert x_m - \gamma_d ^{-1/2} \vert$) is reduced during the transition.
Inversely, planets with marginally librating orbits prior to the disk
depletion may also sometimes attain circulating orbits after the disk
depletion provided the difference between the libration and resonant
centers is increased.

Using these results we infer the kinematic properties of planetary
systems around both Ups And and HD168443.  During the passage of the
secular resonances, the libration time scales of the planets around
these two stars are $2 \times 10^4 $ and $10^5$ years, respectively.
Provided the external axisymmetric disk is depleted on a time scale
longer than these libration time scales, transition through the
secular resonance is adiabatic.

In this contribution, we show how the process of disk depletion may
lead to significant modifications in the eccentricity distribution
within some extrasolar multiple planetary systems.  These changes are
the result of secular perturbation the planets exerted on each other.
But planetary secular perturbation and the axisymmetric potential of
evolving disks alone do not lead to eccentricity excitation. We have
not quantitatively addressed the issue of initial eccentricity
excitation among isolated planets or at least one planet in a multiple
planetary system.  We will discuss these problems in the future.

\acknowledgments MN would like to thank R. Mardling for her helpful
suggestions.  MN used the parallel computer (Silicon Graphics Origin
2000) of Earthquake Information Center of Earthquake Research
Institute, University of Tokyo. This work was partly performed while
MN held a National Research Council Associateship at NASA Ames.  This
work is supported in part by NASA through grant NAG5-10727 and NSF
through AST-9987417 to D. N. C. L..

\appendix

\section{PRECESSION DUE TO THE DISKS' GRAVITY}
\label{sec:disk}

In this section, we evaluate $\Delta N$ in equation (\ref{eq:detadt}).
After protoplanets have acquired a sufficiently large mass, they
induce the formation of a gap near the orbit through tidal interaction
with the disk (Goldreich \& Tremaine 1980; Lin \& Papaloizou 1980,
1993).  While the disk material can no longer be rapidly accreted onto
the protoplanet (Bryden et al. 1999), it continues to contribute to
the gravitational potential.  Although the disk does not have a
sufficiently large mass to modify the angular frequency of the
planet's orbit, its gravity can induce precession (Ward 1981).  The
disk's contribution to the gravitational force in the radial direction
at $r$ is
\begin{equation}
F_r(r) = - {d \Psi \over dr}= - 2 G \int_{0} ^{\infty} y \Sigma (y)
\left( {E(\psi) \over 1 - y} + {K (\psi) \over 1 + y} \right) dy,
\end{equation}
where $\Sigma$ is the surface density of the disk, $\Psi$ is the
potential, $y \equiv r^\prime/r$, $\psi \equiv 2 y^{1/2}/(1+y)$,
$K(\psi)$ and $E(\psi)$ are elliptic integrals of first and second
kind, and $G$ is the gravitational constant (eq. 2-146 in Binney \&
Tremaine 1987).  The associated precession frequencies for planets c
and d can be obtained from
\begin{equation}
N_{c,d} = (\Omega_{c,d} - \kappa_{c,d}) t_c =  {n_{c,d} t_c
\over 2 G M_\ast} \left( {\partial \over \partial r}
r^3 F_r \right)_{r=a_{c,d}},
\label{eq:ncd}
\end{equation}
where $\Omega_{c,d}$ and $\kappa_{c,d}$ are respectively the angular
and epicycle frequencies at semi-major axis of planets c and d
($a_{c,d}$).  Depending on the $\Sigma$ distribution, $N_{c,d} \sim O
(n_{c,d} t_c M_D / M_\ast )$ where $M_D = \Sigma a_{c,d}^2 $ is the
characteristic disk mass at $r=a_{c,d}$ respectively.  From
equation (\ref{eq:ncd}), we find
\begin{equation}
\Delta N= N_c - N_d = {t_c \over 2 G M_\ast} \left(n_c \left({\partial
\over \partial r} r^3 F_r\right)_{r=a_c} -  n_d \left({\partial
\over \partial r} r^3 F_r\right)_{r=a_d} \right),
\label{eq:dncd}
\end{equation}
with its sign determined by the $\Sigma$ distribution.  For disks with
finite $\Sigma$ only outside the planets' orbit, the addition of the
disk self gravity leads to both $N_c$ and $N_d$ to be positive.  Since
planet d is closer to the external disk, $N_d > N_c$ so that $\Delta
N$ is negative.  For disks which engulf both planets' orbits, $\Delta
N$ can still be negative provide $\Sigma$ does not decrease too
rapidly with $r$.  For example, if we use the minimum mass nebula model
(Hayashi et al. 1985) as a fiducial prescription for $\Sigma$ such
that
\begin{equation}
\Sigma = 1.7 \times 10^3 (r/1{\rm AU})^{-1.5} {\rm g} {\rm cm}^{-2},
\label{eq:minebu}
\end{equation}
$\Delta N$ is negative.

\section{ORBITS OF LIBRATING AND CIRCULATING PLANETS}
\label{sec:libra}

Using linearized equations (\ref{eq:dxdtau}) and (\ref{eq:detadt}), we
consider here three families of orbits in systems which contain two
planets.  We assume that the two planets orbit on the same plane as
the disk.

\subsection{The Orbits of Librating Planets}
\label{sec:libra1}

As an example, we consider the planets c and d around Ups And.  For
planets c and d around Ups And, $C= -0.4$, $\gamma_d =0.29$.  The
longitudes of these two planets are aligned such that their $\eta
\simeq 0$ today. The present values of $x=0.66$ and $t_c = 1.1\times
10^3$ years.  With these values, we show below that secular
interaction between the two planets induces $x$ and $\eta$ to undergo
small amplitude oscillation around stationary points.  These points
correspond to extrema for $x$ and $\eta$.  From the linearized
equations (\ref{eq:dxdtau}) and (\ref{eq:detadt}) of secular motion,
we find that these extrema occurs at $\eta=0$ and $\pi$.

For $\eta=0$, a stationary solution can be obtained for
$x=x_m (>0)$ where $x_m$ is a root of the quadratic equation
\begin{equation}
x_m^2 - \left( {\Delta N + 1 - \gamma_d \over \gamma_d C} \right) x_m -
{1 \over \gamma_d} = 0.
\label{eq:xmq}
\end{equation}
With $x=x_m$ and $\eta=0$, longitudes of the periapse of both planets
are aligned and precess together, even though $g_1 \neq g_2$.
Consequently, there is no angular momentum exchange and the
eccentricities of both planets are conserved.  For solutions near the
stationary point with $\eta=\pi$, $x_m$ can be obtained from a similar
equation where
\begin{equation}
x_m^2 + \left( {\Delta N + 1 - \gamma_d \over \gamma_d C} \right) x_m -
{1 \over \gamma_d} =0.
\label{eq:xmqetapi}
\end{equation}

In the ($x$-$\eta$) diagram, these orbits follow the equi Hamiltonian
contours (see Appendix \ref{sec:hamil}).  Angular momentum is being
continually exchanged between the planets, leading to an eccentricity
modulation.  However, the libration is centered around $\eta = 0$ or
$\pm \pi$ where there is no net torque between the two planets.
Although the range of $\eta$ modulation ($\Delta \eta$) is limited,
the two planets are {\it not} in the center of secular resonance
because $\eta$ modulates with a frequency $\omega$ which does not
vanish with arbitrarily small $\Delta \eta$.

We now consider a small perturbation about the stationary point.  For
small values of $\eta$ and $\epsilon \equiv x / x_m - 1$, equations
(\ref{eq:dxdtau}) and (\ref{eq:detadt}) can be linearized in $\eta$
and $\epsilon$ such that
\begin{equation}
{d \epsilon \over d \tau} = {C \over x_m} (1 + \gamma_d x_m^2 ) \eta,
\label{eq:epslin}
\end{equation}
\begin{equation}
{d \eta \over d \tau} = - {C \over x_m} (1 + \gamma_d x_m^2) \epsilon.
\label{eq:etalin}
\end{equation}
Combining these two equations, we find the solution
\begin{equation}
\epsilon = \epsilon_0 {\rm sin} (\omega \tau), \ \ \ \ \ \eta  =
\eta_0 {\rm
cos} (\omega \tau), \ \ \ \ \ {\rm with}
\ \ \ \eta_0 = \pm \epsilon_0,
\label{eq:eta0}
\end{equation}
where the oscillation frequency is
\begin{equation}
\omega = {\pm C  \over x_m}
\left( 1+ \gamma_d x_m^2  \right),
\label{eq:omelin}
\end{equation}
where the signs of the expression for $\eta_0$ and $\omega$ are the
same.  Near the stationary point $x=x_m$ and $\eta=0$, $\omega$ is
also approximately the difference of the precession frequencies of the
two planets because $\varpi$ for both precess in the same direction.

\subsection{Orbit of Circulating Planets}
\label{sec:circu}

As another example, we use the planets around HD168443.  For planets b
and c around HD168443, $C= -0.13$, $\gamma_d =0.14$, and $t_c = 1.9
\times 10^3$ years.  For notational consistency, we use the
subscription for the inner planet is $c$ and that for the outer planet
is $d$.  With these parameters, $\Delta N=0$ for $x_m =0.15$ (or 46)
and $\eta =0$ (or $\pi$). The observed value of $x=2.65$ and $\eta
=1.92$ which is far from the stationary point $x=x_m$ and $\eta=0$ (or
$\pi$).  With these parameters, $\vert A_1 \vert \gg \vert A_2 \vert$
(see eqs. [\ref{eq:rescon}] and [\ref{eq:detadt}]), we expect the
orbits of planets b and c to circulate with respect to each other.  In
this case, the linearized equations (\ref{eq:epslin}) and
(\ref{eq:etalin}) are no longer applicable.  The small magnitude of
$C$ in equation (\ref{eq:dxdtau}) implies that the modulation
amplitude of $x$ is small compared with its average value.  The
dominance of the term of $1-\gamma_d+\Delta N$ in equation
(\ref{eq:detadt}) implies that, to zeroth order in $\tau$, $x \simeq
x_0$ and $\eta \simeq (1 - \gamma_d) \tau$ where $x_0$ is a constant.
Expansion to the next order in $\tau$ yields the solution that
\begin{equation}
\eta \simeq (1 - \gamma_d + \Delta N) \tau + {C \over x_0} {(1 -
\gamma_d x_0^2) \over (1 - \gamma_d + \Delta N) } {\rm sin}
(1 - \gamma_d + \Delta N) \tau,
\label{eq:eta1b}
\end{equation}
\begin{equation}
x \simeq x_0 - {C \over 1-\gamma_d + \Delta N} {\rm cos} (1-\gamma_d
+ \Delta N) \tau,
\label{eq:x1}
\end{equation}
so that the circulation period is $t_l \simeq 2 \pi t_c / (1 - \gamma_d
+ \Delta N) \sim 1.4 \times 10^4$ years for $\Delta N=0$.

\subsection{Non Periodic Orbits Near Secular Resonance}
\label{sec:excite}

In addition to the librating and circulating orbits, there is another
family of non periodic orbits for the case when $\vert A_1 \vert \le
\vert A_2 \vert$.  In the limit that both $x$ and $\eta$ have values
which are very different from $x_m$ and $\eta$=0 (or $\pi$), $\eta$ would
evolve rapidly to a phase angle
\begin{equation}
\eta_1 \simeq {\rm cos}^{-1}  \left( -{A_1 \over A_2} \right)
= {\rm cos}^{-1} \left( {-A_1 x \over C (1 - \gamma_d x^2) } \right),
\label{eq:eta1a}
\end{equation}
such that $d \eta / d \tau $ is, to within the first order
approximation, reduced to zero (see eq. [\ref{eq:detadt}]) and the two
planets become phase locked. Within the same order approximation, $x$
becomes $x_1$ and the evolution of $x$ is determined by
\begin{equation}
{d x \over d \tau} = {d x_1 \over d \tau} \simeq C ( 1 + \gamma_d x_1^2)
{\rm sin} \eta_1 \simeq \pm C ( 1 + \gamma_d x_1^2)
\left( 1 - { A_1^2 x_1^2 \over C ^2 (1 - \gamma_d x_1^2)^2 }
\right)^{1/2},
\label{eq:xcite}
\end{equation}
provided $\vert A_1 \vert$ remains smaller than $\vert A_2 \vert$.
(In Appendix \ref{sec:deplet}, we consider the depletion of the disk
during which the magnitude of $A_1$ evolves with time.)

The plus and minus signs on the right hand side of equation (\ref{eq:xcite})
correspond to $0 \le \eta_1 \le \pi$ and $\pi \le \eta_1 \le 2 \pi$
respectively. Increases/decreases in $x$ are equivalent to
eccentricity excitation/damping for body c.  As a consequence of $x$
evolution, $\eta$ also adjust to $\sim \eta_1$ such that the evolution
of the planets' orbit is non periodic.  Nevertheless, during the
raise/decline in $x$, the two planets may also undergo small amplitude
($x_2, \eta_2$) libration around the evolving guiding center at $x_1$
and $\eta_1$.  The evolution of $x_2$ and $\eta_2$ can be determined
from the next order expansion of equations (\ref{eq:dxdtau}) and
(\ref{eq:detadt}) such that
\begin{equation}
{d x_2 \over d \tau} = B_1 \eta_2 + B_2 x_2,
\label{eq:x2}
\end{equation}
\begin{equation}
{d \eta_2 \over d \tau} = B_o ( \tau - \tau_0) +B_3 x_2 + B_4 \eta_2.
\label{eq:eta2}
\end{equation}
The coefficients are
\begin{eqnarray}
B_0 & = & \Delta {\dot N}, \ \ \ \ \
B_1 = C (1 + \gamma_d x_1^2) {\rm cos} \eta_1, \cr
B_2 & = & 2 C \gamma_d x_1 {\rm sin} \eta_1, \ \ \ \ \ \ \ \ \ B_3
=  - (C/x_1^2) (1 + \gamma_d x_1 ^2) {\rm cos \eta_1}, \cr
B_4 & = & - (C/x_1)(1 - \gamma_d x_1^2) {\rm sin} \eta_1,
\end{eqnarray}
where $\Delta {\dot N} = d \Delta N/d \tau$ 
and the fiducial dimensionless time $\tau_0$ corresponds to the
instance when the two planets first enter into a phase lock, {\it
i.e.}  when their $x=x_1$ and $\eta=\eta_1$. The solution of equations
(\ref{eq:x2}) and (\ref{eq:eta2}) can be expressed as
\begin{eqnarray}
x_2 & = & x_{2 0} {\rm exp} \{\omega_{20} (\tau - \tau_0)\}
+x_{2 1} {\rm exp} \{\omega_{21} (\tau - \tau_0)\}
+ B_5 \tau + B_6, \cr
\eta_2  & = & {1 \over B_1} \left[
(\omega_{20} - B_2) x_{2 0} {\rm exp} \omega_{20} (\tau - \tau_0)
+(\omega_{21} - B_2) x_{2 1} {\rm exp} \omega_{21} (\tau - \tau_0)\right]
\cr
& & \ \ \ \ \ + {1 \over B_1} \left[B_5 ( 1 - B_2 \tau) - B_2 B_6
\right],
\label{eq:solx2}
\end{eqnarray}
where $x_{2 0}$ and $x_{21}$ are the amplitudes of oscillations and
\begin{equation}
B_5  \equiv  {B_0 B_1 \over B_2 B_4 - B_1 B_3} \ \ \ \ \ {\rm and}
\ \ \ \ \ B_6 \equiv {(B_2 + B_4 ) B_5 - B_0 B_1 \tau_0  \over
B_2 B_4 - B_1 B_3},
\label{eq:b5}
\end{equation}
with the growth rate
\begin{equation}
\omega_2 =\omega_{20, 21}= \frac{C}{2x_1}
\left\{ ( 3 \gamma_d x_1^2 -1 ) \sin \eta_1 \ \ \pm \ \ (1 + \gamma_d x_1
^2) (5 \sin^2
\eta_1 - 4)^{1/2}\right\}.
\label{eq:omega2}
\end{equation}
Note that when $\eta_1$ pass through the value ${\rm sin} ^{-1} (4 /
{\sqrt 5})$, $\omega_{20} = \omega_{21}$ and the solution of equations
(\ref{eq:x2}) and (\ref{eq:eta2}) becomes slightly modified from that
in equations (\ref{eq:solx2}).  Also note that for $\eta_1 = 0$ or
$\pi$, so that equations (\ref{eq:x2}) and (\ref{eq:eta2}) reduce to
equations (\ref{eq:epslin}) and (\ref{eq:etalin}) respectively.  Also,
$\omega_2$ becomes purely imaginary ($\pm i \omega$) so all the orbits
undergo libration about the stationary point.  But, for a more general
value of $\eta_1$, $\omega_2$ is complex, and oscillations about the
evolving $x_1$ values either grow or decay depending on the value of
$\eta_1$.  At the center of resonance where $A_1 =0$ and $\eta_1 =
\pi/2$ so that $\omega_2 = 2 C \gamma_d x_1$ or $ (C/x_1) (\gamma_d
x_1^2-1)$.  Both roots are real so that the $x_2$ either converge or
diverge (if $x_1 > \gamma^{-1/2}$) without any oscillation. Note that
oscillation about $x_1$ and $\eta_1$ only occurs if the magnitude of
$\omega_2$ is larger than $B_5$ such that changes in $\Delta N$ and
$x_1$ occur in an adiabatic manner.  For further discussions on the
evolution of planetary orbits during the epoch of disk depletion, see
Appendix \ref{sec:deplet}.

\section{THE EVOLUTION OF MULTIPLE-PLANET SYSTEMS DURING 
THE EPOCH OF DISK DEPLETION}
\label{sec:deplet}

During the depletion of the disk, changes in the magnitude of $\Delta
N$ can cause $A_1$ to change sign.  In \S4, we presented numerical
results of orbital evolution during the epoch of disk depletion.
Here, we present analytical solutions to interpret the numerical
results.  Following the discussions in Appendix \ref{sec:libra}, we
consider three initial orbits separately.

\subsection{The Evolution of Two-Planet Systems with Initially 
Librating Orbits}

In Appendix \ref{sec:libra1}, we show that the condition for libration
is $\vert A_1 \vert \le \vert A_2 \vert$ and $\epsilon_0 \ll 1$. For
planets c and d around Ups And, $\Delta N=0$ today so that $x_m \sim
-C / (1- \gamma_d) \sim 0.5$ and $\omega \sim 0.83$.  Since the
observed values of $x$ and $\eta$ are close to those of $x_m$ and 0
respectively, the small-amplitude oscillator approximation is valid.
The necessary condition for libration, {\it i.e.} $\vert A_1 \vert \le
\vert A_2 \vert$ is marginally satisfied.  From equation 
(\ref{eq:ome}), we find that the time scale for one complete libration
cycle today is $t_l = \tau_{\rm lib} t_c = 2 \pi t_c / \omega = 8.5
\times 10^3$ years.

Prior to its depletion, five times minimum mass nebula model would
lead to $M_D/M_d = \Sigma (a_d) a_d ^2 / M_d \sim 0.015$ and $\Delta N
\sim -1.4$ (eqs. [\ref{eq:dncd}] and [\ref{eq:dnexpsit}]).  In this
case, the necessary condition for libration is also marginally
satisfied and it is possible to find a real positive solution to the
quadratic equation (\ref{eq:xmq}).  Around the stationary point $x_m
\sim 6.5$, there are also solutions which correspond small amplitude
libration between planets c and d ($\vert \eta \vert \lesssim \pi/2 $)
despite the rapid precession induced by the disk's self gravity.  The
corresponding $\omega \sim 0.80$ for such a protoplanetary system is
very similar to the present disk-free value of $x_m$.

During the depletion of the protostellar disk, the magnitude of
$\Delta N$ gradually vanishes so that both $x_m$ and $\omega$ changes
with time.  For $-1.5 < \Delta N < 0$, the condition for libration
($\vert A_1 \vert \le \vert A_2 \vert$) is satisfied.  Provide
$\tau_{\rm lib} < \tau_{A_1}$, $\tau_{\rm lib} < \tau_\omega$,
and $\tau_{\rm lib} < \tau_{x_m}$ where
\begin{equation}
\tau_{\rm lib} \equiv {2 \pi \over \omega} = { 2 \pi x \over C
(1 + \gamma_d x^2) }, \ \ \ \
\tau_{A_1} \equiv {A_1 \over \dot A_1} = { 1 - \gamma_d + \Delta N \over
\Delta \dot N},
\label{eq:taua1}
\end{equation}
\begin{equation}
\tau_{x_m} \equiv {x_m \over \dot x_m} = - \left( { 1 +
\gamma_d x_m ^2 \over 1 - \gamma_d x_m ^2} \right) \tau_{A_1}, \ \ \ \
\tau_\omega \equiv {\omega \over \dot \omega} =
\left( { 1 + \gamma_d x_m ^2 \over 1 - \gamma_d x_m ^2} \right)^2
\tau_{A_1},
\label{eq:tauome}
\end{equation}
the libration of the planets' orbits is preserved while the
librational center and frequency evolves adiabatically.

At the special phase when $A_1 =0$, the second term in equations
(\ref{eq:xmq}) and (\ref{eq:xmqetapi}) reduces to zero and the
stationary points with $\eta=0$ and $\eta=\pi$ coincide with each
other with $x_m= \gamma_d^{-1/2}$.  We refer to this special
configuration as librational degeneracy.  In this state, all orbits
librate either around $x_m = \gamma_d ^{-1/2}$, $\eta =0$ or $\pi$
with two separatrix located on the $\eta = \pi/2$ and $3 \pi/2$ (see
Hamiltonian contours maps).  From equation (\ref{eq:dg12}), we note that in
this degenerate state, $\Delta g_{12}$ attains a minimum value with
respect to variations in $\Delta N$.  The oscillation frequency around
the stationary points $\omega$ attains a minimum magnitude $2 \gamma_d
^{1/2} C$ with $x_m= \gamma_d^{-1/2}$ (see eq. [\ref{eq:ome}]) which is
$\sim 1.8$ for planets c and d around Ups And.

Note that in this state of librational degeneracy, $g_1 \neq g_2$ in
general.  However, if planet c has negligible mass, $\Delta g_{12}$
would vanish and it would enter into a state of exact secular
resonance and $x_m$ would diverge.  This asymptotic state represent
the eccentricity excitation of asteroids by the Jovian planets.  For
finite values of $\gamma_d$, this state of librational degeneracy can
also be considered as a state of secular resonance despite the finite
value of $\Delta g_{12}$.  The secular interaction between the planets
forces them to librate so that their relative longitude of periapse
passage is confined to a limited amplitude.

We now consider the evolution of the librational amplitude.  The
action associated with the simple oscillator equations
(\ref{eq:epslin}) and (\ref{eq:etalin})
\begin{equation}
J = \int \dot \epsilon d \epsilon = \epsilon_0 ^2 \omega \int_0
^{2 \pi} {\rm cos}^2 \theta d \theta = \pi \epsilon_0 ^2 \omega
\label{eq:jacti}
\end{equation}
is invariant to first order in $\tau_{\rm lib} / \tau_{A_1}$ or
$\tau_{\rm lib} / \tau_\omega$ (Goldstein 1980), which ever is larger.
(For values of $\tau_{A_1}$ and $\tau_\omega$, see eqs. [\ref{eq:taua1}]
and [\ref{eq:tauome}]).  Based on this invariance, we can also determine
the changes in $\epsilon_0$ which is $\propto \omega^{-1/2}$.  For
planets c and d around Ups And, we adopted a model in which $\omega
\sim 0.8$ both during the embedded phases and after the disk is
completely depleted.  Thus, we expect both $\epsilon_0$ and $\eta_0$
to be remain invariant despite large changes in $x_m$.  The numerical
results in \S4.1 is in agreement with this analytic derivation.

But, the condition for adiabatic evolution may not always be satisfied
especially when the two planets pass through a phase of secular
resonance with $x_m = \gamma_d ^{-1/2}$ when the libration period
takes the maximum value $t_l = -\pi t_c /(C \gamma_d^{1/2})= 1.7\times
10^4$ years and $\omega$ attains a minimum value of $ \pm 2 C
\gamma^{1/2} = \mp 0.43$ (compared with its initial value of $\sim
0.8$).  During this phase, the libration time scale $\tau_{\rm lib}$
reaches a maximum. Although at the center of secular resonance,
$\tau_{A_1}$ vanishes with $A_1$, the more appropriate value for it is
$\sim \Delta N/ \Delta \dot N$ which remains finite.  Nevertheless, it
is possible for $\tau_{\rm lib}$ to be larger than $\tau_{A_1}$.  Note
that $\tau_{x_m}$ reduces to $\simeq \pm 2 / \Delta \dot N$ in the
resonance.  In accordance with equation (\ref{eq:tauome}),
$\tau_\omega$ diverges.  But, the characteristic time scale
for $\omega$ evolution based on a second order expansion remains
finite.  The numerical results in \S4.1 show a slight increase in
$\epsilon_0$ with a $\sim 6 \%$ reduction in $\eta_0$.  In contrast, a
40 \% increase in $\epsilon_0$ is inferred from equation 
(\ref{eq:jacti}).  Nevertheless, for planets c and d around Ups And,
the duration of non adiabatic evolution is brief so that the small
amplitude libration around $x=x_m$ and $\eta=0$ is preserved despite
the large net changes in the magnitude of $x_m$.

Note that we have only consider the axisymmetric contribution of the
disk to the total potential.  Therefore the total angular momentum of
the two planets is conserved during the disk depletion even though
there is a net transfer of angular momentum between them which leads
to the decline in $e_c$ and the growth in $e_d$.

\subsection{The Evolution of Two-Planet Systems with
Initially Circulating Orbits 1: Asteroids}

We now consider the orbital evolution of multiple-planet with initial
values of $\vert A_1 \vert \ge \vert A_2 \vert$.  In Appendix
\ref{sec:circu}, we showed that these systems have initially
circulating orbits.  There are two possible branches of solutions.
First, we consider the limiting case of small $\gamma_d$ ($\ll
x^{-2}$) which is appropriate in the context of secular interaction
between the asteroids and Jupiter for which $M_c \ll M_d$.

During the epoch of disk depletion, as the magnitude of $A_1$ in
equation (\ref{eq:detadt}) is reduced below that of $A_2$, $\eta$ evolves to
$\eta_1$ (see eq. [\ref{eq:eta1a}]) which reduces to $\eta_1 = {\rm
cos}^{-1} (-A_1 x /C)$ in the limit $\gamma_d \ll x^{-2}$.  Since
$A_1$, $A_2$, and $C$ are negative, $\eta_1 = \pi$ when the planets
first enter into a secular resonance.  Also, equation (\ref{eq:xcite})
implies that $x_1$ decreases/increases for $\eta_1$ slightly
less/greater then $\pi$.

Let us first consider the case that $\eta_1$ is slightly less than
$\pi$. Continual decline in $x$ is attainable provided $d \eta_1 / d
\tau \le 0$.  The reduction in $x$ leads to an increase in the
magnitude of $A_2$ while the disk depletion causes a decrease in that
of $A_1$.  Consequently, the magnitude of $A_1/A_2$ decreases in
equation (\ref{eq:eta1a}), leading to a reduction in $\eta_1$.  This
state of phase lock can only be maintained if $\eta$ can adjust to
$\eta_1$ and decrease to $\sim \pi/2$. So long this state of resonant
phase locking is maintained, the magnitude of $x$ ({\it i.e.} $x_1$)
can decrease indefinitely.  When $A_1$ passes through the zero value
and becomes positive, $x$ continues to decline.  If the magnitude of
$x$ cannot be reduced as fast as that of $\Delta N$, $\vert A_1 \vert
> \vert A_2 \vert$ and the planets would attain circulating orbits.
Even in the limit that the magnitude of $x$ can be reduced as fast as
that of $\Delta N$, the planets are unlikely to attain librating
orbits because their $\eta_1 \sim \pi/2$. Eventually, $\eta_1$
terminates its evolution with values slightly less than $\pi/2$ while
the orbits of the massless particles are approximately circularized.

If the planets enter into the secular resonance with $\eta$ slightly
larger than $\pi$, $x$ would increase.  Further excitation of $x$
requires $d \eta_1 / d \tau \ge 0$.  In equation (\ref{eq:detadt}), the
growth in $x$ reduces the magnitude of $A_2$ while that of $A_1$ is
also decreasing.  Phase lock can only be maintained if $\vert A_1
\vert < \vert A_2 \vert$ such that $x$ cannot grow faster than $C/A_1
\simeq C/(1 - \Delta N)$ during the epoch of disk depletion.  If $x$
grows slower than $C/A_1$, $d \eta / d \tau >0$ and $\eta_1$ would
increase to enhance the $x$ growth rate.  The magnitude of $\eta_1$ is
determined by the ratio of $A_1$ and $A_2$ (see eq. [\ref{eq:eta1a}]).
The evolution time scale of $A_2$ equals to $\tau_{x_m}$.
Equations (\ref{eq:taua1}) and (\ref{eq:tauome}) imply that $\tau_{A_1}
\simeq \tau_{x_m}$ in the limit of small $\gamma_d x^2$. Thus,
magnitude of $A_1/A_2$ is regulated to be less than unity as phase
lock causes $x$ to increase (see eq. [\ref{eq:xcite}]).

As the resonant condition $1 - \Delta N \simeq 0$ is approached,
$\eta_1$ increases to $3 \pi/2$ and $x$ grows at a maximum rate $d x /
d \tau = -C$.  So long this state of resonant phase locking is
maintained, the magnitude of $x$ can increase indefinitely.  On the
resonance, $g_1 = g_2$ such that the growth occurs regardless of the
value $\eta$.  Increases in $x$ is equivalent to eccentricity
excitation for the less massive body c.  When $A_1$ passes through the
zero value and becomes positive, $x$ can have a sufficiently large
value such that $e_c$ may approach and even exceed unity.  This
eccentricity excitation process has been noted by Heppenheimer (1980)
as a potential cause for the existence of highly eccentric asteroids.

As $x$ increases in the resonances, there are three eventual outcomes:
a) $x$ becomes sufficiently large for $e_c$ to exceed unity and body c
to escape, b) the planets would attain circulating orbits as the
magnitude of $A_1$ exceeds that of $A_2$ after $\vert \Delta N \vert$
decreases below $\sim 1$ if $\tau_{A_1}$ becomes much smaller than
$\tau_{x_1}$, and c) in systems with finite $\gamma_d$, the above
approximation become inappropriate when $x$ exceeds $\gamma_d^{-1/2}$
(see below).

\subsection{The Evolution of Two-Planet Systems with Initially 
Circulating Orbits 2: Two Planets with Comparable Masses}

We now consider the second branch of solutions in which $\gamma_d$ is
finite.  These parameters are more appropriate for the extrasolar
multiple-planet systems in which the masses of the planets are
comparable. The main difference introduced by the finite value of
$\gamma_d$ is the possibility of a transition as the value of $x$
passes through $\gamma_d ^{-1/2}$.  This branch of solution is
applicable for planets b and c around HD168443.  We use the analytic
solutions in this subsection to study the numerical results in \S4.

For planets b and c around HD168443, $C= -0.13$, $\gamma_d =0.13$, and
$t_c = 1.9 \times 10^3$ years.  In this system, the value of $x_m$
depends sensitively on $\Delta N$ due to the small magnitude of $C$
and $\gamma_d$. Within the uncertainty of analytic estimate, we
estimate $M_D/M_d \sim 0.007$ and $\Delta N \simeq -4$ for five times minimum
mass nebula model.  Despite this relatively small value of $M_D (< M_d
< M_\ast)$, the corresponding $x_m \simeq 180$.  At this initial stage,
$\vert A_1 \vert > \vert A_2 \vert$ and the planets follow circulating
orbits which is well approximated by equations (\ref{eq:eta1b}) and
(\ref{eq:x1}).

But, during the depletion of the disk, the decline of $\Sigma$ reduces
the magnitude of $\Delta N$ which passes through a stage of secular
resonance in which $\Delta N \simeq \gamma_d -1 = -0.86$ with the
corresponding value of $x_m \simeq \gamma_d ^{-1/2} = 2.6$ similar to
the present value of $x$. In this limit, $\vert A_1 \vert < \vert A_2
\vert$ and the leading-order approximation with the solutions in
equations (\ref{eq:eta1b}) and (\ref{eq:x1}) are no longer valid.
Since the planets' orbits circulate relative to each other prior to
entering this stage, the small-oscillation-amplitude approximation
which led to the solutions in equation (\ref{eq:eta0}) may also be
inappropriate.  Nevertheless, when the planets first enter into the
resonance, $\vert A_1 \vert = \vert A_2 \vert$, $\eta_1 = \pi$, and
$\omega_2$ in equation (\ref{eq:omega2}) becomes purely imaginary. The
discussions in \S\ref{sec:excite} indicates that during the passage
through the resonance, all orbits are librating regardless of the
amplitude of the libration cycles.

Within the resonance where $\vert A_1 \vert < \vert A_2 \vert$, the
two planets become phase locked such that $x_1$ and $\eta_1$ evolve in
accordance to equations (\ref{eq:eta1a}) and (\ref{eq:xcite}).
Depending on the rate of disk depletion, there are three potential
outcomes for passage through the secular resonance. 1) In the limit
that $\tau_{A_1} < \tau_{\rm lib}$, the system evolves in an impulsive
manner.  Such a situation is relevant for rapid clearing of disks
which contain long-period, low-mass, widely-separated planets with
large $t_c$.  In this case, there is insufficient time for the
longitude of periapse to significantly circulate before the effect of
secular resonance fades with the rapid disk depletion. During this
rapid passage through the secular resonance, both $x$ and $\eta$
essentially retain the values they had when they first entered into
the resonance ($x_1$ and $\eta_1$).  These values of $\eta_1$ and
$x_1$ become the initial values of circulating orbits in the new
disk-free potential.  Since $\eta_1$ and $x_1$ are random parameters,
the actual variational amplitude of the circulating orbits are
unpredictable.

2) For systems such as planets b and c around HD168443, $\tau_{A_1} <
\tau_{\rm lib}$ so that the planets adiabatically evolve into a phase
lock.  The relevant governing equations become equations 
(\ref{eq:eta1a}) and (\ref{eq:xcite}). In contract to the case for the
massless asteroids, the system first enters or finally exits the
secular resonances with $\eta =0$ if $x > \gamma_d^{-1/2}$ or $\pi$ if
$x < \gamma_d^{-1/2}$.  In the middle of the resonant zone where
$\gamma_d -1 - \Delta N =0$, although $x_m =\gamma_d^{-1/2}$, $x$
generally does not equal $x_m$ so that $\eta =\pm \pi/2$. (Even in the
limit that $x = x_m$ at the center of resonances, $A_1/A_2$ also
vanishes at the center of the resonance). Our analysis in \S
\ref{sec:excite} indicates that, at the center of resonance, both
roots of $\omega_2 (= 2 C \gamma_d x_1$ or $ (C/x_1) (1 - \gamma_d
x_1^2)$) are real so the amplitude of small perturbations either
converge or diverge (if $x_1 > \gamma_d^{-1/2}$) from the stationary
points without any oscillation. Note that in contrast to the case of
massless asteroids, $\omega_2$ changes sign as $x_1$ passes through
$\gamma_d ^{-1/2}$ so the system can neither be damped nor be excited
indefinitely.  This sign reversal of $\omega_2$ limits the among of
changes in $x_1$. Physically, this limit is due to the comparable
momentum of inertia carried by planets of similar masses.  In such
systems, the exchange of angular momentum cannot lead to vastly
different eccentricity changes for the two planets.

After passing through the center of the resonance, $\eta_1$ continues
to evolve towards 0 or $\pi$ in accordance to equation (\ref{eq:eta1a})
while $x_1$ evolves in accordance to equation (\ref{eq:xcite}).  When,
$\eta_1$ reduces below sin$^{-1}$0.8, $\omega_1$ once again become
complex and libration around $x_1$ and $\eta_1$ would resume if the
magnitude of $\omega_2$ is larger than $B_5$ such that changes in
$\Delta N$ and $x_1$ occur in an adiabatic manner.  But in equation 
(\ref{eq:b5}), the magnitude of $B_5$ is determined by $\Delta \dot N$
or equivalently $\tau_{A_1}$.  Thus, in the modestly rapid depletion
case where $\tau_{A_1} \le \omega_2^{-1}$, the difference between
$x_1$ and $x_m$ at the center of resonance is preserved.  For small C
magnitude (as in the case of HD168442), the width of the libration is
small and $x_m$ is also not changed significantly during the passage
through the resonance.  Thus, planets would not be able to enter into
a librating state and the circulation pattern of their orbits is
retained.

3) Finally, in the slow depletion limit where $t_{A_1} > \omega_2^{-1}
$, the protracted passage through the secular resonance can greatly
modify both $x$ and $\eta$.  As the planets exit the resonance, they
can become trapped onto librating orbits.  Large changes in $x$
implies that at least one planet attains a relatively large
eccentricity. In relatively compact multiple-planet systems, large
eccentricity can cause mean motion resonances to overlap and dynamical
instability.  We shall address the issue of dynamical instability
elsewhere.

\section{HAMILTONIAN CONTOURS}
\label{sec:hamil}

These phase space evolution of the planets' orbits may be compared
with equi-Hamiltonian contour maps.  By expanding the disk potential
in equation (\ref{eq:vpot}) for planets with eccentric orbits and then
averaging it over their orbital period, Ward (1981) obtained the
disturbing function due to the disk potential as follows:
\begin{equation}
<R_{\rm disk}>_{c, d}=e_{c, d}^2\cdot\pi G \Sigma_0 a_{c,d}
\left(\frac{a_0}{a_{c,d}}\right)^{k}\sum_{n=1}^{\infty}A_n\frac{n(2n
+1)} {2n+k-1} \left( \frac{a_{c, d}}{r_{\rm edge}}\right)^{2n+k-1} \equiv
e_{c,d}^2 T_{c,d}.
\label{eq:upsnebdis}
\end{equation}
Using this potential, $\Delta N$ of equation (\ref{eq:dncd}) can be
written as
\begin{equation}
\Delta N= \frac{8a_d^2}{G a_c b_{3/2}^{(1)}}\left(T_c-T_d
\left( \frac{a_c}{a_d}\right)^{1/2} \right).
\label{eq:dnexpsit}
\end{equation}
When the disk depletion time scale is longer than characteristic
librational time scale associated with equations (\ref{eq:dxdtau}) and
(\ref{eq:detadt}) (time scales are given in Appendix
\ref{sec:deplet}), $\Delta N$ can be regarded as quasi static.  In
this limit, the planets' orbits follow a single trajectory on
($x$-$\eta$) plane depending on the magnitude of their Hamiltonian
and angular momentum.  Unless the orbits of the interacting planets
are locked in mean motion resonances, the orbital energy and
semi-major axes of each planet are conserved to the lowest order.  The
orbit-averaged Hamiltonian ($H$) can be calculated semi-analytically
using the disturbing functions of planets (e.g., Brouwer \& Clemence
1961) and equation (\ref{eq:upsnebdis}). Expanding to the second order
of the eccentricities, we find
\begin{equation}
H=\frac{M_c n_c^2 a_c^2}{2}+\frac{M_d n_d^2 a_d^2}{2}
+\frac{n_c a_c^2 N_c}{2t_c}+\frac{n_d a_d^2 N_d}{2t_c}
+\frac{GM_cM_da_c}{8a_d^2}
\left\{b_{3/2}^{(1)}(e_c^2+e_d^2)-2b_{3/2}^{(2)
}e_ce_d\cos\eta \right\}.
\end{equation}
From the conservation of the total angular momentum, a relation
\begin{equation}
J \equiv M_{c}n_{c}a_{c}^2\sqrt{1-e_{c}^2}+
M_{d}n_{d}a_{d}^2\sqrt{1-e_{d}^2}={\rm constant  }
\label{eq:ang}
\end{equation}
holds provided their orbital planes are not inclined from each other.
Thus, to second order in eccentricity,
\begin{equation}
\frac{n_c a_c^2 N_c}{t_c}+\frac{n_d a_d^2
N_d}{t_c}+\frac{GM_cM_da_c}{4a_d^2}
\left\{b_{3/2}^{(1)}(e_c^2+e_d^2)-2b_{3/2}^{(2)
}e_ce_d\cos\eta \right\}={\rm constant}, \label{eq:hamicon}
\end{equation}
\begin{equation}
M_{c}n_{c}a_{c}^2e_{c}^2+M_{d}n_{d}a_{d}^2e_{d}^2={\rm constant},
\label{eq:angcon}
\end{equation}
for any given values of $H$ and $J$.  Divided equation 
(\ref{eq:hamicon}) with equation (\ref{eq:angcon}), we find that
\begin{equation}
\frac{x^2(1+N_c)+2 C x \cos\eta +(\gamma_d^{-1}+N_d)}{\gamma_d x^2+1}={\rm
constant},
\label{eq:equihami}
\end{equation}
and obtain the equi-Hamiltonian contour lines for various values of
the constants of integration.  In equation (\ref{eq:equihami}), the
masses of the disk and planets are contained only in the form of their
ratios.  Therefore, as long as $M_{\ast}\gg M_{c, d}$, the
equi-Hamiltonian contours are not affected by the obliquity $i$ of the
system.  We show below that other than small deviations due to
non-linear effects, the planets' orbits closely follow the contours in
the equi-Hamiltonian map.

\clearpage
\begin{figure}
\plotone{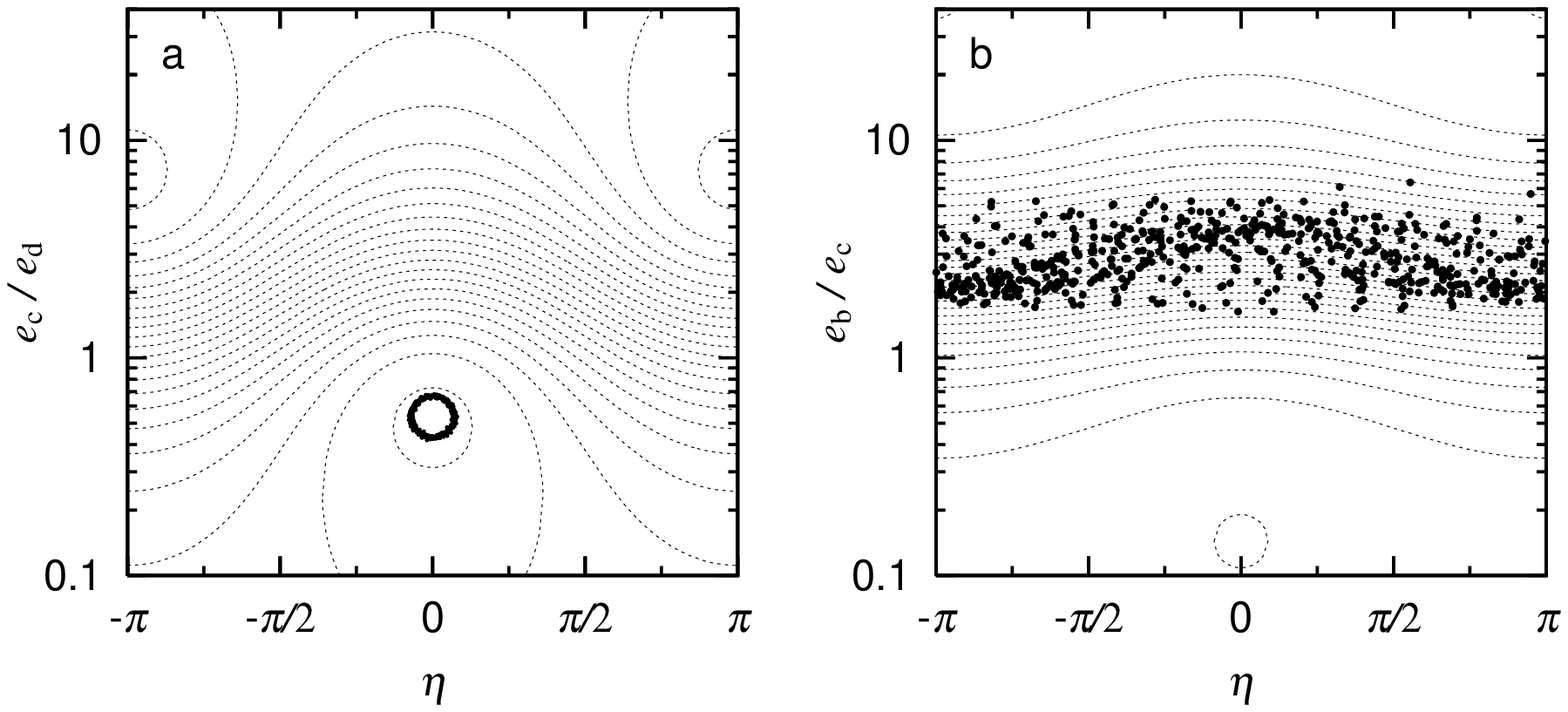}
\caption{Orbital evolutions of extra solar planets.  Points show
the orbits obtained from numerical simulations.  The dotted lines are
the equi-Hamiltonian contours of the planetary system.  The contours
are spaced with equal intervals.  (a)Orbital evolution of planets
c and d (points) of Upsilon Andromedae system.  Orbits are integrated
for $2\times 10^4$ years starting from present orbital elements.  The
period of the libration is about 6000 years.  (b)Orbital
evolution of planets b and c (points) of HD168443 system.  Orbits are
integrated for $2.5\times10^4$ years. The circulation period for $\eta$
to evolve from $-\pi$ to $\pi$ is about 16000 years.
\label{fig:orbitnow}
}
\end{figure}

\clearpage
\begin{figure}
\plotone{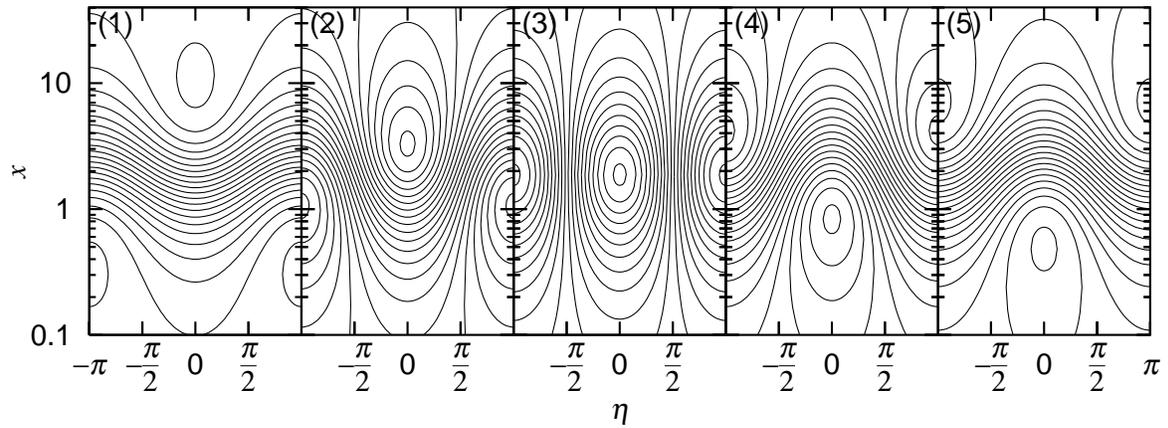}
\caption{Equi-Hamiltonian contours of Upsilon Andromeda planetary
system in ($x$-$\eta$) diagram.  The time sequence of nebula depletion
proceeds from panel (1) to (5).  The nebula edge is set to be (1)
4.2AU, (2) 5AU, (3) 5.5AU, (4) 7AU, and (5) infinity (the present
disk-free planetary system).
\label{fig:upshmp}
}
\end{figure}

\clearpage
\begin{figure}
\plotone{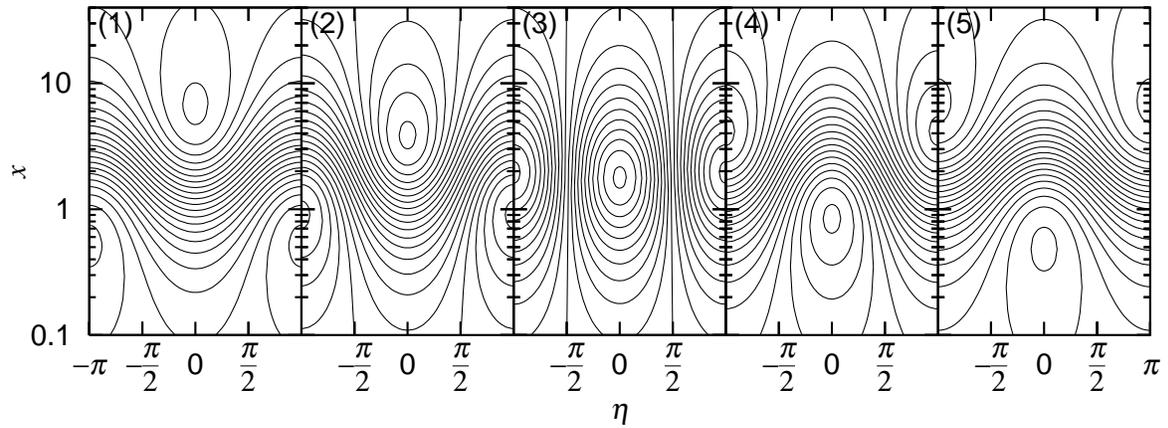}
\caption{Equi-Hamiltonian contours of Upsilon Andromeda planetary
system in ($x$-$\eta$) diagram.  The time sequence of nebula depletion
proceeds from panel (1) to (5).  The depletion ratio
$\exp(-t/t_{\Delta N})$ is at (1)1, (2) 0.75, (3) 0.5, (4) 0.25, and
(5) 0. The disk edge is located at 4.5AU.
\label{fig:hmpupsuni}
}
\end{figure}

\clearpage
\begin{figure}
\plotone{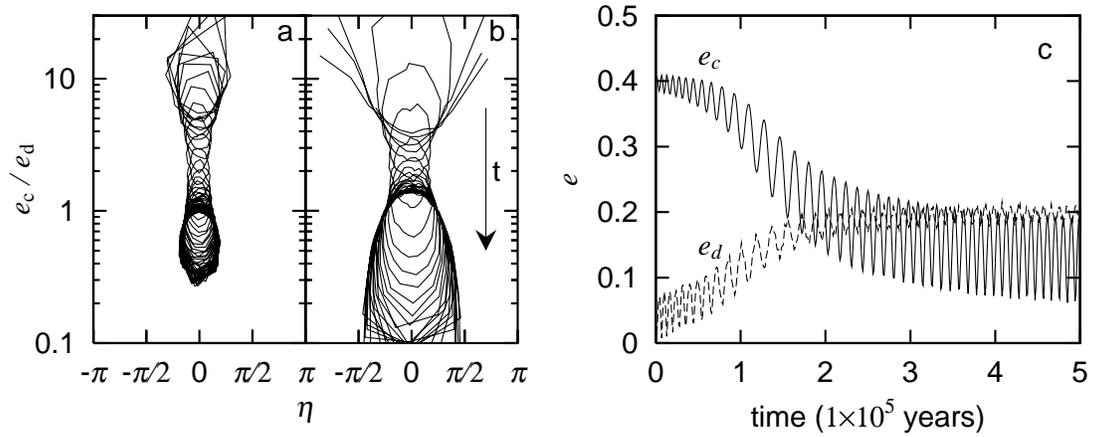}
\caption{Orbital evolutions of planets during the depletion of the
protoplanetary disk.  (a) In this system, planets were initially on a
librating track in the ($x$-$\eta$) diagram and remained on a librating
track with a high $e_{d}$ after their nascent disk is completely
depleted. The initial magnitude of $x=8$ and $\eta =\pi/4$ prior to
the epoch of the disk depletion.  (b) In this system, planets were
initially on a circulating track in the ($x$-$\eta$) diagram but entered on
a librating track after the disk is completely depleted. The initial
magnitude of $e_{c}=0.4$, $e_{d}=0.05$, and $\eta=3\pi/2$ prior to the
disk depletion.  (c) The evolution of the eccentricities in the case
of (a). The solid and dashed lines show the eccentricity of planet c
and d, respectively.  The retreating speed of nebula edge is
$10^{-5}$(AU ${\rm year}^{-1}$), and the nebula edge is 4.2 AU at $t=0$.
\label{fig:upseome}
}
\end{figure}

\clearpage
\begin{figure}
\plotone{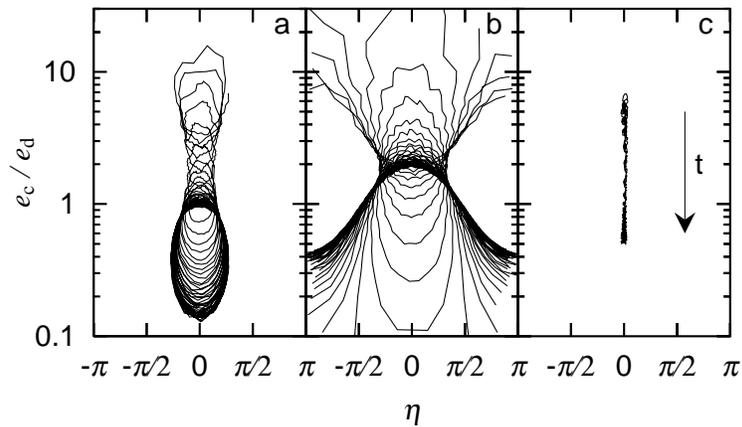}
\caption{The orbital evolutions of planets during the uniform
depletion of the protoplanetary disk.  (a) In this system, planets
were initially on a librating orbit.  The initial condition is the
same as the case of Figure \ref{fig:upseome}a.  (b) In this system,
planets were initially on an open circulating track and remained on an
open circulating track after the disk is completely depleted. The
initial magnitude of $e_{c}=0.4$, $e_{d}=0.05$, and $\eta=\pi$.  (c)
In this system, the coincidence of the periastrons is preserved.  The
magnitude of $e_{c}=0.45$, $e_{d}=0.08$, and $\eta=0$ initially.  In
these calculations, the nebula edge is set to be 4.5 AU, and
$t_{\Delta N}=10^5$ years.
\label{fig:upseomeuni}
}
\end{figure}

\clearpage
\begin{figure}
\plotone{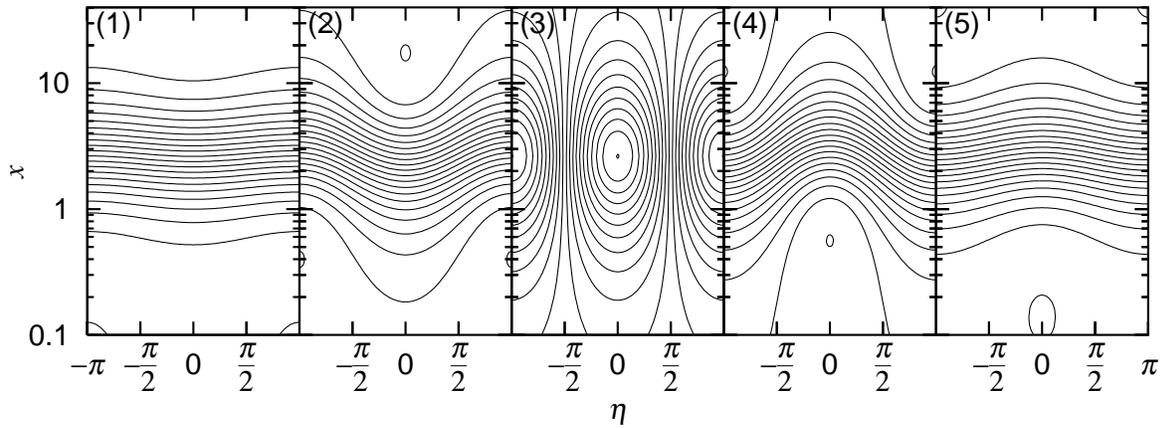}
\caption{Equi-Hamiltonian contours of HD168443 planetary system in the
($x$-$\eta$) diagram.  The time sequence of the nebula depletion
proceeds from panel (1) to (5).  The nebula edge is at (1) 5AU, (2)
6.5AU, (3) 7.2AU (secular resonance) (4) 8AU, and (5) infinity
(present planetary system).
\label{fig:hdhmp}
}
\end{figure}

\clearpage
\begin{figure}
\plotone{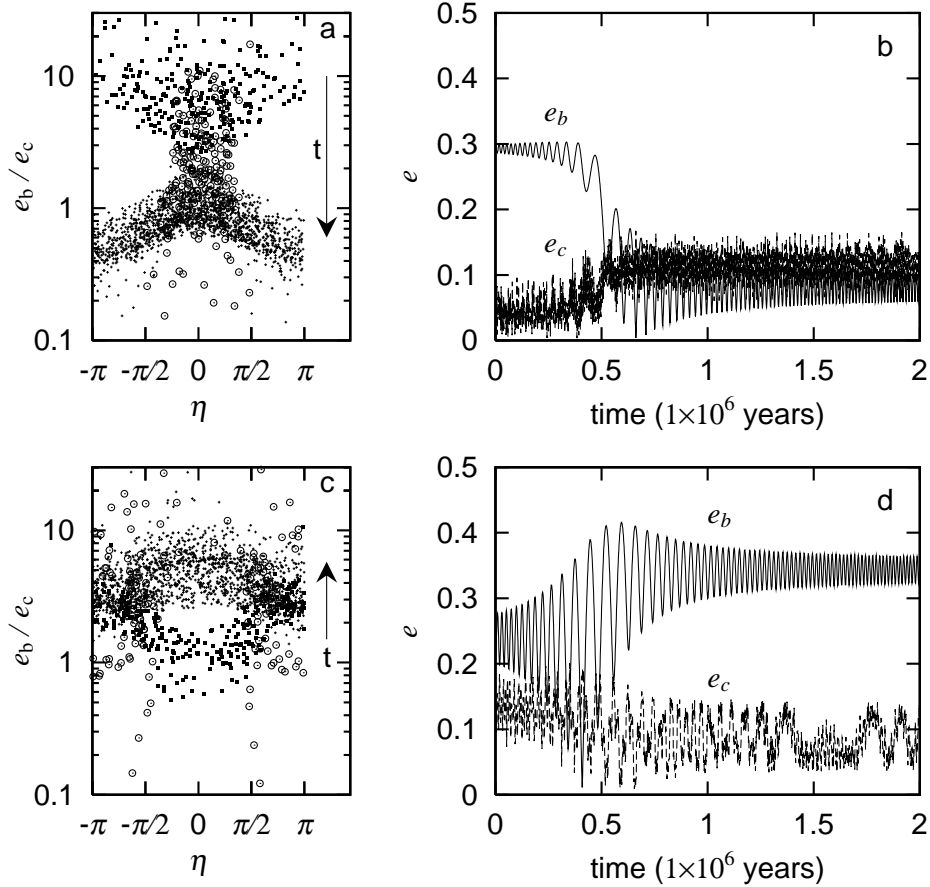}
\caption{Orbital evolutions of HD168443 planets during the uniform
depletion of the protoplanetary disk. The edge of the disk is located
at $r_{\rm edge}=5.5$AU.  The disk depletion timescale is $t_{\Delta
N}=10^6$ years.  (a) The orbital evolution in the ($x$-$\eta$) diagram
for the case with the initial values of $x=6$ and $\eta=0$. The
eccentricity ratio for the first $t= 4\times 10^5$ years is denoted by
filled squares and that for $t=4\times 10^5\sim 7\times 10^5$ years
(around the secular resonance) is denoted by open circles and that for
$t>7\times 10^5$ years are shown by small plus symbols.  (b) The
evolution of eccentricities corresponds to the system of planets in
panel a.  (c) The evolution of the eccentricity ratio for the case
with the initial values of $x=1$ and $\eta=0$.  Filled squares, open
circles, and small plus show the period of time before $4\times 10^5$,
during $4\times 10^5\sim 6.2\times 10^5$, and after $6.2\times 10^5$
years, respectively.
\label{fig:hduni}
}
\end{figure}

\clearpage
\begin{figure}
\plottwo{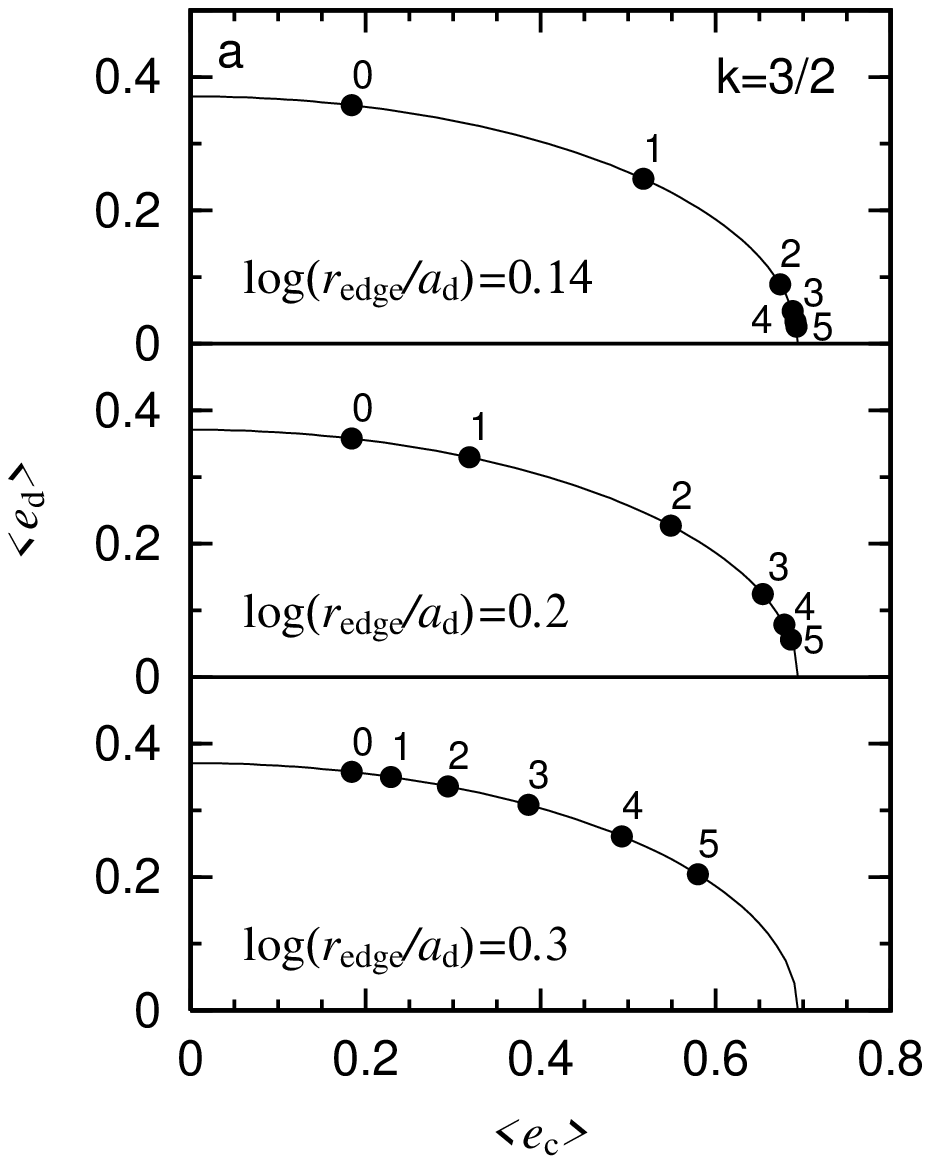}{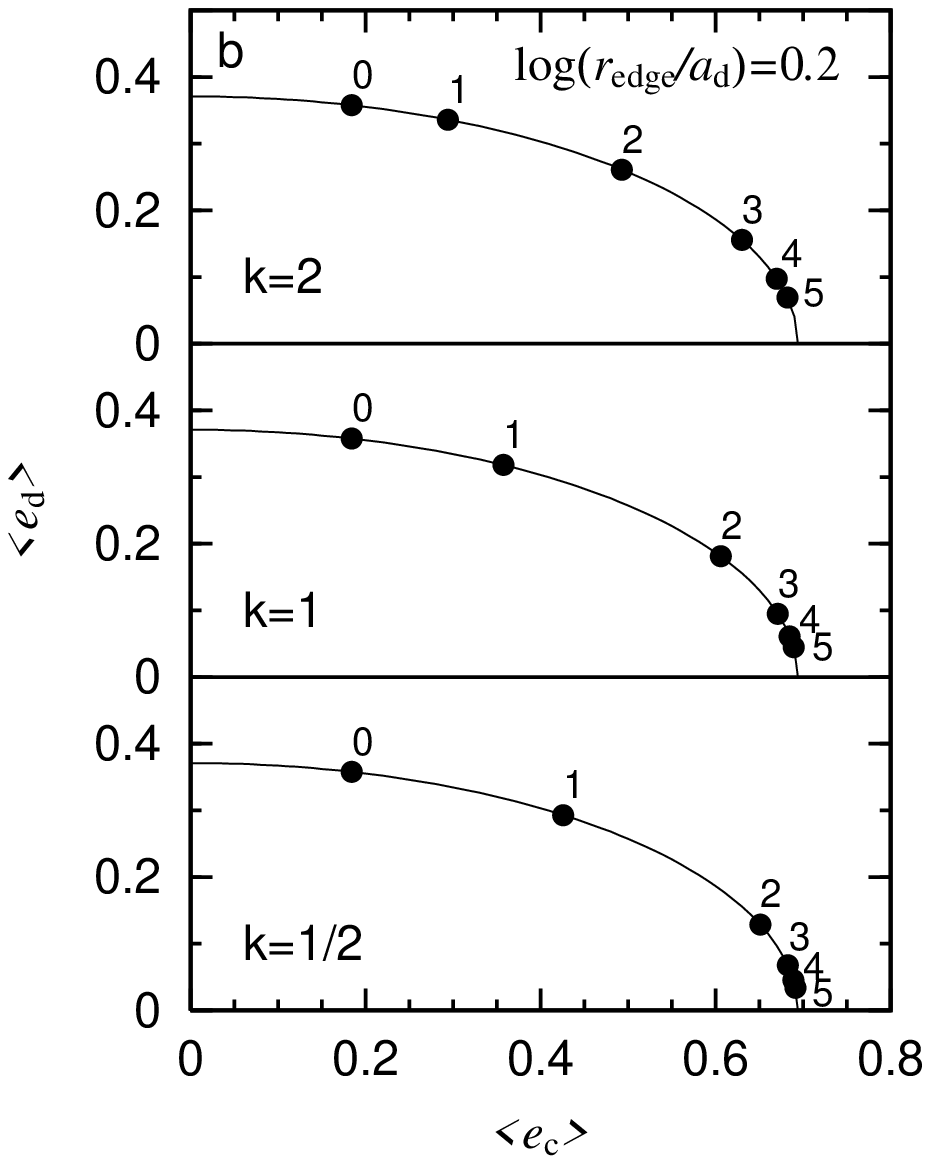}
\caption{Initial eccentricities of planets c and d around Ups And.
The eccentricities $e_c$ and $e_d$ are shown by filled circle for five
deferent initial disk mass.
The number along the filled circle shows the surface density (e.g., '5'
means $5\sin i \times$ of the minimum mass model).
(a)The locations of the disk edge are $\log(r_{\rm edge}/a_d)=0.14$ (top
panel), $0.2$ (middle panel), $0.3$ (bottom panel).
(b) The surface density gradients are $k=3/2$ (top panel), $k=1$ (middle
panel), $k=1/2$ (bottom panel). The location of the disk edge is 4.057AU.
\label{fig:modeldep}
}
\end{figure}

\clearpage
\begin{figure}
\plotone{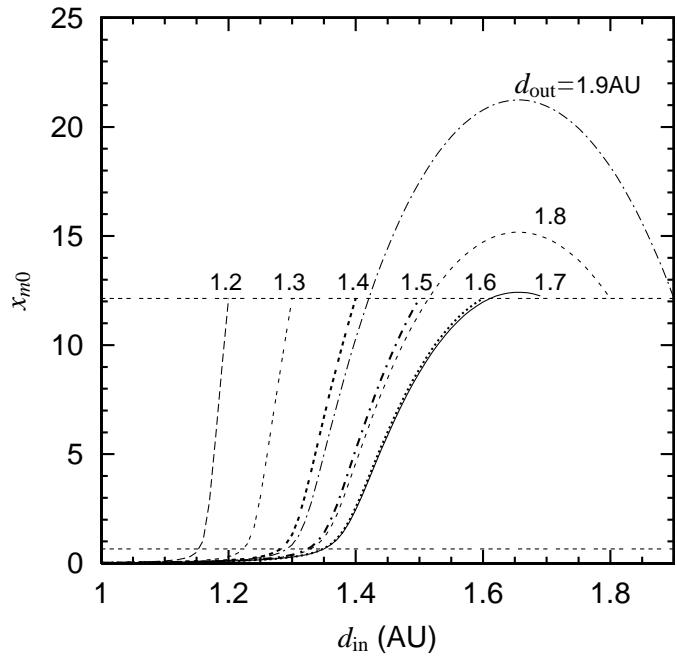}
\caption{The location of the librating center around zero ($x_{m0}$)
in the case that there is a ring between the planets c and d around
Ups And. The ring's surface density is set to be $5 \sin i$ times that of the 
minimum  mass model and it is confined between $d_{\rm in}$ and $d_{\rm out}$. 
Eight lines for different values of $d_{\rm out}$ are shown. The
horizontal line at $\sim 12$ shows the $x_{m0}$ without the disk. The
horizontal line at 0.66 shows the current $x_{m0}$.
\label{fig:ring}
}
\end{figure}

\clearpage
\begin{figure}
\plotone{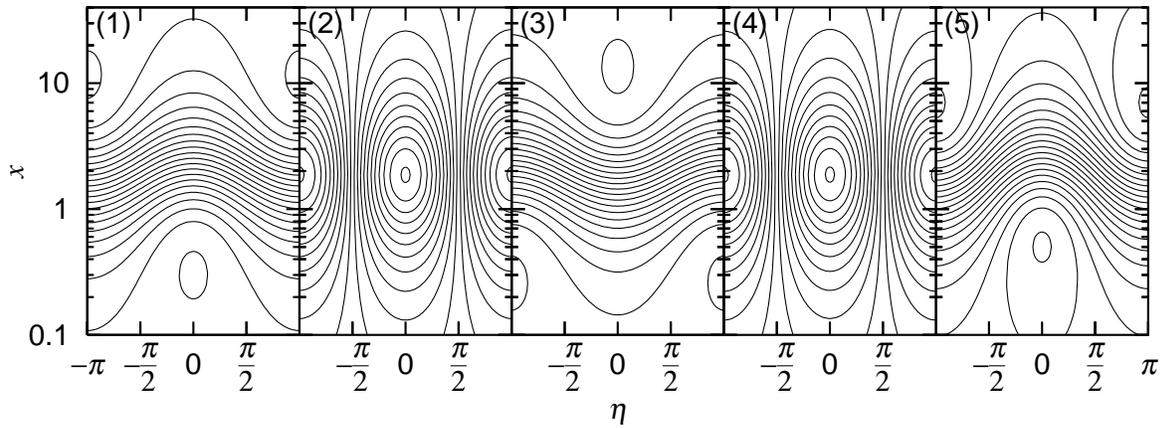}
\caption{Equi-Hamiltonian contours of Ups And planetary system.  The
surface densities of residual ring and the disk are (1)100\% , (2)
53\% (ring) and 100\% (disk), (3) 0\% (ring) and 100\% (disk), (4) 0\%
(ring) and 35\% (disk), and (5)0\% of their initial values,
respectively.
\label{fig:ringhmp}
}
\end{figure}

\clearpage
\begin{figure}
\plotone{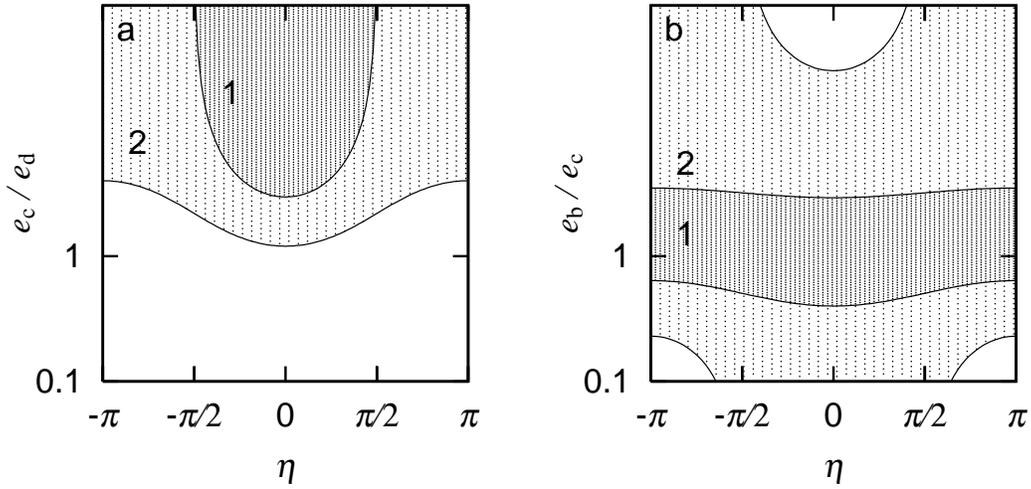}
\caption{(a)A schematic diagram of the ($x$-$\eta$) plane for the
system of planets around Ups And.  Planets which initially occupy
region 1, enter into a librating track in the ($x$-$\eta$) plane after
the disk depletion.  Their orbits become similar to present orbits of
planets c and d.  For planets which initially occupy region 2, their
eccentricity ratio is reduced after the disk depletion.  (b) A
schematic diagram of the ($x$-$\eta)$ plane for the planetary systems
around HD168443.  Planets which initially occupy region 2, enter into
the open circulating tracks in the ($x$-$\eta$) plane after the disk
depletion.  Planets which initially occupy region 1 evolve into orbits
similar to those observed around HD168443.
\label{fig:upsang}
}
\end{figure}

\clearpage
\begin{deluxetable}{ccccccccl}
\tablecolumns{9}
\tablewidth{0pc}
\tablecaption{Orbital parameters of extrasolar multiple planets
planets}
\tablehead{
\colhead{planets}& \colhead{}&
\colhead{$a$(AU)}& \colhead{}&
\colhead{$e$}&\colhead{}&
\colhead{$\varpi$(rad)}&
\colhead{$M\sin i$($M_{J}$)}&
\colhead{$T_{\rm peri}$}
}
\startdata
\sidehead{(a) Upsilon Andromedae ($M_{\ast}=1.3 M_{\rm \odot}$)}
planet b  && 0.059 && 0.01 && 5.522  & 0.69 & 2450000.6383 \\
planet c  && 0.827 && 0.23 && 4.314  & 2.06 & 2450399.0    \\
planet d  && 2.56  && 0.35 && 4.374  & 4.10 & 2451348.9    \\
\tableline
\sidehead{(b) HD168443 ($M_{\ast}=1.01 M_{\rm \odot}$)}
planet b  && 0.295 && 0.53 && 3.018 & 7.73  & 2450047.58 \\
planet c  && 2.87  && 0.20 && 1.098 & 17.15 & 2450250.6 \\
\tableline
\sidehead{(c) HD74156 ($M_{\ast}=1.05 M_{\rm \odot}$)}
planet b  && 0.276 && 0.649 && 3.206 & 1.56  & 2451981.4 \\
planet c  && 3.47  && 0.395 && 4.189 & $>$7.5 & 2450849 \\
\tableline
\sidehead{(d) 47 UMa ($M_{\ast}=1.03 M_{\rm \odot}$)}
planet b  && 2.09 && 0.061 $\pm$0.014 && 3.00 $\pm$0.26 & 2.54  &
2453622$\pm$34 \\
planet c  && 3.73  && 0.1$\pm$0.1 && 2.22 $\pm$0.98 & 0.76 & 2451363.15
$\pm$493\\
\tableline
\sidehead{(e) HD12661 ($M_{\ast}=1.07 M_{\rm \odot}$)}
planet b  && 0.804 && 0.35 && 5.11 & 2.21  & 2460208.393 \\
planet c  && 2.652  && 0.11 && 2.29 & 1.58 & 2459398.08
\enddata \label{tab:parameta}
\tablerefs{http://exoplanets.org/almamacframe.html}
\end{deluxetable}

\clearpage
\begin{deluxetable}{cccc}
\tablecolumns{4}
\tablewidth{0pc}
\tablecaption{Notations}
\tablehead{
\colhead{Symbol}& \colhead{meaning}
}
\startdata
$C$ & $-5a_c/(4a_d)$ \\
$\gamma_{d}$ & $(a_c/a_d)^{1/2}(M_c/M_d)$\\
$\epsilon$ & $x/x_m-1$\\
$\eta$ & $\varpi_c-\varpi_d$ \\
$t_c$ & $(4 /3 n_c)(M_\ast / M_d) (a_d/a_c)^3$\\
$x$ & $e_c/e_d$ \\
$x_m$ & values of $x$ at centers of librating orbits\\
$\Delta N$ & $d\eta/d\tau$ caused by the disk potential (eq. [\ref{eq:dncd}])
\enddata \label{tab:variables}
\end{deluxetable}
\clearpage

\end{document}